\documentclass{article}
\usepackage[margin=1in]{geometry}
\usepackage{amsmath}
\usepackage{epsfig}
\usepackage{tcolorbox}
\numberwithin{equation}{section} 
\newcommand{\beq}{\begin{equation}}
\newcommand{\eeq}{\end{equation}}
\newcommand{\beqa}{\begin{eqnarray}}
\newcommand{\eeqa}{\end{eqnarray}}
\newcommand{\bdm}{\begin{displaymath}}
\newcommand{\edm}{\end{displaymath}}
\newcommand{\lslash}[1]{#1\llap/}

\newcommand{\Eq}[1]{Eq.\ (\ref{#1})}
\newcommand{\Eqs}[2]{Eqs.\ (\ref{#1}) and (\ref{#2})}

\newcommand{\xRef}[1]{Ref.\ \cite{#1}}
\newcommand{\Fig}[1]{Fig.\ \ref{#1}}
\newcommand{\Table}[1]{Table\ \ref{#1}}
\newcommand{\Tr}{\mbox{Tr}\,}
\newcommand{\Section}[1]{Section\ \ref{#1}}
\newcommand{\Appendix}[1]{Appendix\ \ref{#1}}
\newcommand{\fEp}{f}
\newcommand{\fEpp}{f^\prime}
\newcommand{\fbEp}{\bar f}
\newcommand{\fbEpp}{\bar f^\prime}
\newcommand{\fnu}{f^\prime_\nu}
\newcommand{\fnub}{\bar f^\prime_\nu}

\title{
  Neutrino decoherence in an electron and nucleon background
}
  
\author{Jos\'e F. Nieves\footnote{nieves@ltp.uprrp.edu}\\
  Laboratory of Theoretical Physics, Department of Physics\\
  University of Puerto Rico, R\'{\i}o Piedras, Puerto Rico 00936
  \and\\[12pt]
  Sarira Sahu\footnote{sarira@nucleares.unam.mx}\\
  Instituto de Ciencias Nucleares\\
  Universidad Nacional Aut\'onoma de Mexico\\
  Circuito Exterior, C. U.\\
  A. Postal 70-543, 04510 Mexico DF, Mexico\\
}
\date{}
\begin{document}
\maketitle

\begin{abstract}
  We consider the decoherence effects in the propagation of active neutrinos
  due to the non-forward neutrino scattering processes
  in a matter background composed of electrons and nucleons.
  We calculate the contribution to the imaginary part of the neutrino
  self-energy arising from such processes. Since the initial neutrino state is
  depleted but does not actually disappear (the initial neutrino
  transitions into a neutrino of a different flavor but does not decay)
  those processes should be associated with decoherence effects
  that cannot be described in terms of the coherent evolution
  of the state vector. Based on the formalism developed in our previous work
  for treating the non-forward scattering processes using
  the notion of the stochastic evolution of the state,
  we identify the jump operators, as used in the context of the master or
  Lindblad equation, in terms of the results of the the calculation
  of the non-forward neutrino scattering contribution to
  the imaginary part of the neutrino self-energy.
  As a guide to estimating the decoherence effects in situations 
  of practical interest we  give explicit
  formulas for the decoherence terms for different background conditions,
  and point out some of the salient features in particular the
  neutrino energy dependence.
  To establish contact wih previous works in which the decoherence
  terms are treated as phenomenological parameters, we consider
  the solution to the evolution equation in the two-generation case.
  We give formulas that are useful for estimating
  the effects of the decoherence terms under various conditions
  and environments, including the typical conditions applicable
  to long baseline experiments, where matter effects are important.
  In those contexts the effects appear to be small, and indicative that
  if significant decoherence effects were to be found
  they would be due to non-standard contributions to the
  decoherence terms.
\end{abstract}

\section{Introduction and Summary}
\label{sec:introduction}

It is well known that neutrinos propagating through a background medium
acquire an index of refraction produced by their coherent,
forward scattering, interaction processes with the background particles.
One approach is to calculate the real (or dispersive)
part of the neutrino self-energy in the context of
Thermal Field Theory (TFT)\cite{ftft:reviews},
from which the neutrino and antineutrino effective potential and dispersion
relations can be determined\cite{Nieves:2018vxl}.

The neutrino interactions with the background particles can also
produce damping terms in the neutrino effective potential and
index of refraction. In a previous work\cite{nsnuphidamp} we considered
the calculation of such damping terms in a background of fermions ($f$)
and scalars ($\phi$) as a consequence of processes such as
$\nu + \phi \leftrightarrow f$ and $\nu + \bar f\leftrightarrow \bar\phi$,
involving the coupling of neutrinos to those particles
of the generic form $\bar f_R\nu_L\phi$. There we
calculated the imaginary part (or more precisely the absorptive part)
of the neutrino self-energy, from which the damping terms in the effective
potential and dispersion relation were obtained.

Subsequently we pointed out that, in addition to the damping effects,
those couplings induce decoherence effects in the propagation of neutrinos
due to the neutrino non-forward scattering process\cite{nsnuphidecoherence}.
More precisely, we considered various neutrino flavors ($\nu_{La}$)
interacting with a scalar and fermion with a coupling of the form
\beq
\label{Lfnuphia}
L_{int} = \sum_a \lambda_a\bar f_R \nu_{La} \phi + h.c.
\eeq
The scattering processes of the form
$\nu_a + x \rightarrow \nu_b + x$, where $x = f,\phi$,
can induce decoherence effects in the propagation of neutrinos,
independently of the possible damping effects mentioned above.
Our strategy there was to determine the contribution of such processes
to the absorptive part of the self-energy, from which we obtained
the corresponding contribution to the damping matrix $\Gamma$ by
the usual method. However, in the case considered there, in which the initial
neutrino state is depleted but does not actually disappear
(the initial neutrino transitions into a neutrino of a different flavor
but does not decay into a $f\phi$ pair, for example), we pointed out that
the effects of the non-forward scattering
processes are more properly interpreted in terms of decoherence phenomena
rather than damping. Thus, we gave a precise prescription to identify the
decoherence terms, specifically the \emph{jump} operators ($L_n$)
as used in the context of the master or Lindblad
equation\cite{Daley:2014fha,Weinberg:2011jg,
pearle,Plenio:1997ep,Lieu:2019cev}, in terms of the results
of the calculation of the imaginary part of the neutrino
self-energy due to the non-forward neutrino scattering processes.
As usual, the formulas for the jump operators
involve integrals over the momentum distribution
functions of the background particles, and as a guide to estimating
such decoherence effects, the relevant quantities
were computed explicitly in the context of the model we considered,
for several limiting cases of the momentum distribution functions
of the background particles.

As a follow-up of that work on the contribution of non-forward
scattering processes to the decoherence effects on the propagation
of neutrinos in a thermal background, here we consider the case
of the standard interactions of neutrinos
with a matter (electron and nucleon) background.
This is of course a realistic situation rather than a hypothetical model,
with potentially important consequences for many research activities
of current interest, from both theoretical and experimental perspectives.

Decoherence effects, in the framework of open systems or
the Lindblad equation, have been considered in the recent neutrino physics
literature in a variety of contexts\cite{lisi,farzan,%
oliveira:hepph160308065,Oliveira:2014jsa,Carpio:2017nui},
and in specific settings such as IceCube\cite{Coloma:2018idr},
DUNE\cite{Gomes:2018inp}, and long base line
experiments\cite{Coelho:2017byq,Gomes:2020muc}. It has also been
considered for their possible relevance in connection
with quantum gravitational effects\cite{Fogli:2007tx}, and
the question of $CPT$ symmetry and the Dirac vs Majorana nature of
neutrinos\cite{Capolupo:2018hrp,Buoninfante:2020iyr}.
Some of these works have explored the dependence of
the decoherence terms on the neutrino
energy (e.g., Refs.\cite{lisi,farzan,Coloma:2018idr,Gomes:2020muc}),
but they have been based on general considerations at a phenomenological level
of the decoherence terms, without a precise calculation of them.

Our work is complementary to this line of work in the sense
that our focus is the calculation of the decoherence terms,
or more precisely the jump operators,
and in this work we concentrate on the case that they
arise from the Standard Model interaction of the neutrinos
with the background particles of the medium in which they propagate.
Our main goal is a precise prescription to determine them
as used in the context of the master or Lindblad equation,
from the calculation of the non-forward neutrino scattering
contribution to the imaginary part of the neutrino self-energy.
The result is a well-defined formula for the decoherence terms in
that context, expressed in terms of integrals over the background
matter fermion distribution functions and standard model couplings of the
neutrino with the electron and nucleons. To establish contact with
the previous works cited, we consider the solution to the evolution equation
in the two-generation case, and we evaluate explicitly the
decoherence parameters for different background conditions
and point out some of their salient features, such
as their neutrino energy dependence once the background conditions are
specified.

\begin{figure}
\begin{center}
\epsfig{file=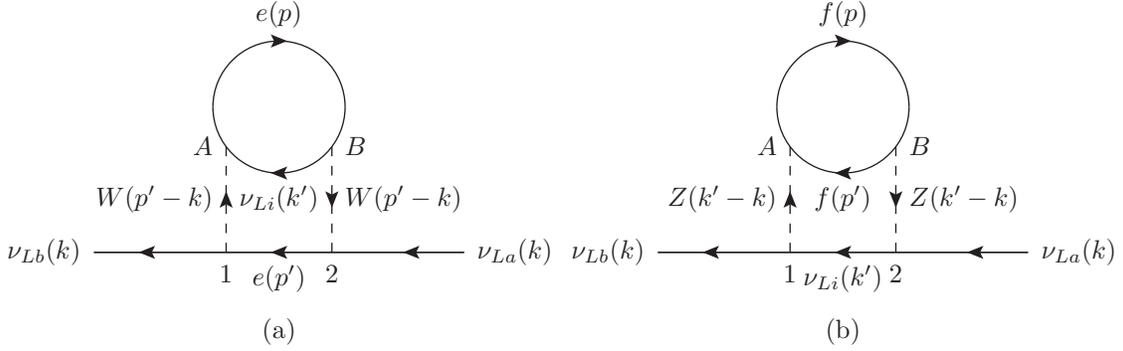,bbllx=93,bblly=342,bburx=510,bbury=466}
\end{center}
  \caption{
  Two-loop diagrams for the damping term in the
  neutrino thermal self-energy in a matter (electron and nucleon) background.
  In Diagram (b) the label $f$ stands for either $e,n,p$.
  In principle we have to consider the various thermal vertices $A = 1,2$
  and $B = 1,2$. However, in the heavy $W,Z$ limit, only the
  diagonal components of the $W,Z$ thermal propagators are non-zero
  and therefore only one diagram, with $A = 1$ and $B = 2$, must be considered.
  For simplicity of notation, we have labeled $k^\prime = p - p^\prime + k$.
  \label{fig:smgeneric}
}
\end{figure}

\begin{figure}
\begin{center}
\epsfig{file=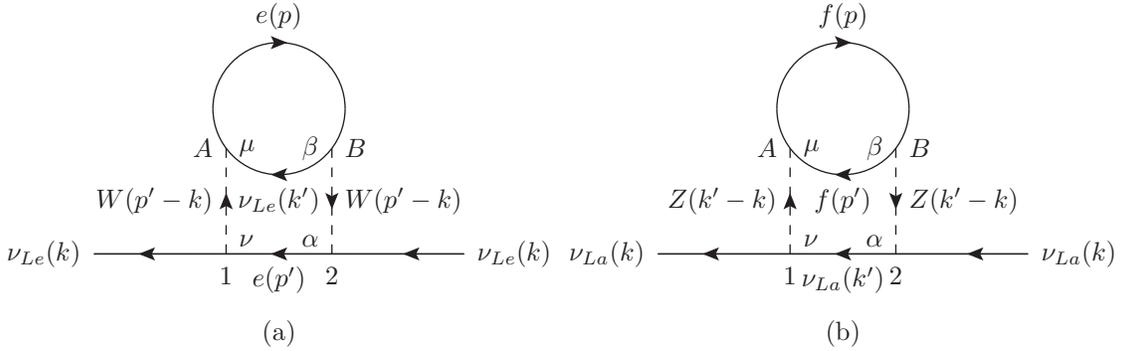,bbllx=93,bblly=342,bburx=510,bbury=466}
\end{center}
\caption{
  Same diagrams as in \Fig{fig:smgeneric} but restricted to the case
  of standard neutrino couplings in the standard model.
  In diagram (a) only $\nu_e$ can participate, while
  in diagram (b) the neutrino line labeled $\nu_a$ can be anyone of the
  flavors $\nu_{e,\mu,\tau}$, but it only contributes to the flavor-diagonal
  matrix element. The rest of the symbols have the same
  meaning as in \Fig{fig:smgeneric}.
  \label{fig:sm}
}
\end{figure}

The diagrams that contribute to the decoherence effects
that we are considering are displayed in \Fig{fig:smgeneric}.
In those diagrams we are labeling the neutrino lines in a generic way, leaving
open the possibility that the active neutrinos may have
non-standard couplings and/or may mix with non-standard (sterile) neutrinos,
for example. But in our calculations for definiteness
we will restrict ourselves to the case of active neutrinos with
standard couplings and mixings, in which case the diagrams
are labeled as shown in \Fig{fig:sm}.

In \Section{sec:preliminaries} we review briefly
our strategy to determine the jump operators
from the results of the calculation
of the absorptive part of the self-energy. This material is based on our
previous work\cite{nsnuphidecoherence}, and we therefore limit
ourselves there to state the main points omitting some details.
In \Section{sec:Gamma} we proceed to the actual calculation
as outlined in \Section{sec:preliminaries}. The end result is a set
of formulas for the decoherence terms, expressed as integrals over
the distribution functions of the background particles.
In \Section{sec:twogenexample} we consider the solution to the evolution
equation in the two-generation case with the decoherence terms we have
obtained, making contact with the previous in which the decoherence
terms are treated as phenomenological parameters.
In \Section{sec:evaluationintegrals} we evaluate explicitly the integrals
involved for some specific simple cases of the background conditions,
which serve as a guide to practical applications.
We use those results in \Section{sec:discussion} to give explicit
formulas for the decoherence parameters in various environments of
potential interest. We give special attention to the typical conditions
applicable to long baseline experiments, where the effects appear to be small,
and indicative that if significant decoherence effects were to be found
they would be due to non-standard contributions to the
decoherence terms. \Section{sec:conclusions} contains our conclusions
and we give in two appendices some of the details of the calculations.
\section{Preliminaries}
\label{sec:preliminaries}

\subsection{Self-energy and the damping matrix}
\label{sec:dampingmatrix}

The following material is borrowed from \xRef{nsnuphidecoherence},
which we summarize here for completeness.
We denote by $u^\mu$ the velocity four-vector of the
background medium and by $k^\mu$ the momentum of the
propagating neutrino. In the background medium's own rest frame,
\beq
\label{restframe}
u^\mu = (1,\vec 0)\,,
\eeq
and in this frame we also write
\beq
\label{krestframe}
k^\mu = (\omega,\vec\kappa)\,.
\eeq
Since we consider only one background medium, which
can be taken to be at rest, we adopt \Eqs{restframe}{krestframe}
throughout.

Let us consider first the case of one neutrino propagating
in the medium, ignoring flavor mixing.
The dispersion relation $\omega(\vec\kappa)$ and the spinor of
the propagating mode are determined by solving the equation
\beq
\label{eveq}
\left(\lslash{k} - \Sigma_{eff}\right)\psi_L(k) = 0\,,
\eeq
where $\Sigma_{eff}$ is the neutrino thermal self-energy,
which can be decomposed in the form
\beq
\Sigma_{eff} = \Sigma_r + i\Sigma_i\,,
\eeq
where $\Sigma_r$ is the dispersive part and $\Sigma_i$ the absorptive part.
In the context of thermal field theory,
\beq
\Sigma_r = \Sigma_{11r} \equiv
\frac{1}{2}(\Sigma_{11} + \overline\Sigma_{11})\,,
\eeq
where $\Sigma_{11}$ is the $11$ element of the thermal self-energy matrix.
On the other hand, $\Sigma_i$ is conveniently obtained from the formula
\beq
\label{Sigmaidef1}
\Sigma_i = \frac{\Sigma_{12}}{2i n_F(x_\nu)}\,,
\eeq
where $\Sigma_{12}(k)$ is the $12$ element of the neutrino
thermal self-energy matrix, while
\beq
\label{nzfermion}
n_F(z) = \frac{1}{e^z + 1}\,,
\eeq
is the fermion distribution function, written in terms of a dummy variable $z$,
and the variable $x_\nu$ is given by
\beq
x_\nu = \beta k\cdot u - \alpha_\nu\,.
\eeq
$\Sigma_{11}$ and $\Sigma_{12}$ will be determined
by evaluating the diagrams shown in \Fig{fig:sm}.

The chirality of the interactions imply that\cite{footnote1}
\beq
\label{SigmaV}
\Sigma_{eff} = V^\mu\gamma_\mu L\,,
\eeq
and correspondingly
\beq
\label{SigmariVri}
\Sigma_{r,i} = V^\mu_{r,i}\gamma_\mu L\,,
\eeq
with
\beq
\label{Vri}
V^\mu = V^\mu_r + iV^\mu_i\,.
\eeq
In general both $V^\mu_{r,i}$ are functions of $\omega$ and $\vec\kappa$.
Ordinarily we will omit those arguments but we will restore them when needed.

Writing the neutrino and antineutrino dispersion relations in the form
\beq
\label{disprelform}
\omega^{(\nu,\bar\nu)} = \omega^{(\nu,\bar\nu)}_r -
\frac{i\gamma^{(\nu,\bar\nu)}}{2}\,,
\eeq
$\omega^{(\nu,\bar\nu)}_r$ is given by
\beq
\label{nudisprelreal}
\omega^{(\nu,\bar\nu)}_r = \kappa + V^{(\nu,\bar\nu)}_{eff}
\eeq
where $V^{(\nu,\bar\nu)}_{eff}$ are the effective potentials
\beqa
\label{Veff}
V^{(\nu)}_{eff} & = & n\cdot V_r(\kappa,\vec\kappa) =
 V^0_r(\kappa,\vec\kappa) - \hat\kappa\cdot\vec V_r(\kappa,\vec\kappa)
\,,\nonumber\\
V^{(\bar\nu)}_{eff} & = & -n\cdot V_r(-\kappa,-\vec\kappa) =
 -V^0_r(-\kappa,-\vec\kappa) + \hat\kappa\cdot\vec V_r(-\kappa,-\vec\kappa)\,,
\eeqa
with
\beq
\label{nmu}
n^\mu = (1,\hat\kappa)\,.
\eeq
On the other hand, for the imaginary part,
\beqa
\label{nudisprelimg}
-\frac{\gamma^{(\nu)}(\vec\kappa)}{2} & = & 
\frac{n\cdot V_i(\kappa,\vec\kappa)}
{1 - n\cdot\left.
\frac{\partial V_r(\omega,\vec\kappa)}{\partial\omega}
\right|_{\omega = \kappa}}\,,\nonumber\\
-\frac{\gamma^{(\bar\nu)}(\vec\kappa)}{2} & = & 
\frac{n\cdot V_i(-\kappa,-\vec\kappa)}
{1 - n\cdot\left.
\frac{\partial V_r(\omega,-\vec\kappa)}{\partial\omega}
\right|_{\omega = -\kappa}}\,,
\eeqa
where $n^\mu$ is defined in \Eq{nmu}.
If the correction due to the
$n\cdot\partial V_r(\omega,\vec\kappa)/\partial\omega$ in the denominator
can be neglected, the formulas in \Eq{nudisprelimg} reduce to
\beqa
\label{nudisprelimg-simple}
-\frac{\gamma^{(\nu)}(\vec\kappa)}{2} & = & 
n\cdot V_i(\kappa,\vec\kappa)\,,\nonumber\\
-\frac{\gamma^{(\bar\nu)}(\vec\kappa)}{2} & = & 
n\cdot V_i(-\kappa,-\vec\kappa)\,.
\eeqa

When we consider various neutrino flavors, the vector $V^\mu$ 
defined through \Eq{SigmaV} is a matrix in neutrino flavor space.
Then, as shown in \xRef{nsnuphidecoherence},
generalization of the discussion above is that the dispersion
relations of the propagating modes are determined by solving
the following eigenvalue equation, in flavor-space,
\beq
\left(H_r - i\frac{\Gamma}{2}\right)\xi = \omega\xi\,,
\eeq
with $H_r$ and $\Gamma$ being Hermitian matrices in flavor space,
calculated in terms of the vector $V_\mu$,
\beqa
\label{HrGammaVrelation}
H_r & = & \left\{\begin{array}{ll}
\kappa + n\cdot V_r(\kappa,\vec\kappa) & (\nu)\\
\kappa - n\cdot V^\ast_r(-\kappa,-\vec\kappa) & (\bar\nu)
\end{array}\right.\nonumber\\[12pt]
-\frac{1}{2}\Gamma & = & \left\{\begin{array}{ll}
n\cdot V_i(\kappa,\vec\kappa) & (\nu)\\
n\cdot V^\ast_i(-\kappa,-\vec\kappa) & (\bar\nu)
\end{array}\right.\,.
\eeqa
In coordinate space, this translates to the evolution equation
\beq
\label{eveqt}
i\partial_t\xi(t) = \left(H_r - i\frac{\Gamma}{2}\right)\xi(t)\,.
\eeq

Our purpose is to determine the contribution to $\Gamma$
due to the diagrams in \Fig{fig:sm}. Our strategy is first to
determine the loop-expression for $\Sigma_{i}$, which follows
from the corresponding loop-expression for $\Sigma_{12}$ by means of
\Eq{Sigmaidef1}. Then use the fact that the
corresponding expression for $V^\mu_i$ is obtained by substituting
the loop-expression for $\Sigma_{i}$ in the formula
\beq
\label{ViSigmailoopformula}
V^\mu_i = \frac{1}{2}\Tr\gamma^\mu \Sigma_i\,,
\eeq
as implied by \Eq{SigmariVri}, which allows to calculate $\Gamma$
by means of \Eq{HrGammaVrelation}. Specifically, we will denote by
$\Sigma^{(W)}_{i}$ the contribution from diagram (a)in \Fig{fig:sm} 
and by $\Sigma^{(Z,f)}_{i}$ the
contribution from diagram (b) for any of the fermions $f = e,n,p$, so that
\beq
\Sigma_i = \Sigma^{(W)}_i + \sum_{f = e,n,p} \Sigma^{(Z,f)}_i\,.
\eeq
From \Eqs{HrGammaVrelation}{ViSigmailoopformula} we then obtain
the loop formula for the damping matrix
\beqa
\label{GammaVfinal}
-\frac{1}{2}\Gamma^{(\nu)} & = &
n\cdot V^{(W)}_i(\kappa,\vec\kappa) +
\sum_{f = e,n,p} n\cdot V^{(Z,f)}_i(\kappa,\vec\kappa)\,,\nonumber\\
-\frac{1}{2}\Gamma^{(\bar\nu)} & = & 
n\cdot V^{(W)\ast}_i(-\kappa,-\vec\kappa) +
\sum_{f = e,n,p} n\cdot V^{(Z,f)\ast}_i(-\kappa,-\vec\kappa)\,.
\eeqa
for neutrinos and antineutrinos, respectively, where
\beqa
\label{ViZfSigmailoopformula}
V^{(Z,f)\alpha}_i & = & \frac{1}{2}\Tr\gamma^\alpha \Sigma^{(Z,f)}_i\,,
\nonumber\\
V^{(W)\alpha}_i & = & \frac{1}{2}\Tr\gamma^\alpha \Sigma^{(W)}_i\,,
\eeqa

\subsection{Jump operators}
\label{sec:decoherence}

Similarly to the case discussed in \xRef{nsnuphidecoherence},
the damping matrix in the present case,
calculated from \Fig{fig:sm} as we have outlined
above, arises from the non-forward neutrino scattering processes,
and not from neutrino decay processes. In this case the initial
neutrino state is depleted but does not actually disappear
and, as we argued, the damping matrix should be associated with decoherence
effects in terms of the Lindblad equation
and the notion of the stochastic evolution of the state
vector\cite{Daley:2014fha,Weinberg:2011jg, pearle,Plenio:1997ep,Lieu:2019cev}.
The idea is to assume that the evolution due to the
damping effects described by $\Gamma$ is accompanied by a stochastic evolution
that cannot be described by the coherent evolution of the state vector.
As discussed in detail in \xRef{nsnuphidecoherence} but
omitting the details here, the result of this idea is that
the evolution of the system in this case is described by the density
matrix (in the sense that we can use it to calculate averages of quantum
expectation values) that satisfies the Lindblad equation,
\beq
\partial_t\rho = -i[H_r,\rho] + \sum_n
\left\{L_n \rho L^\dagger_n - \frac{1}{2}L^\dagger_n L_n\rho -
\frac{1}{2}\rho L^\dagger_n L_n\right\}\,,
\eeq
where the $L_n$ matrices, representing the jump operators,
are related to $\Gamma$ by
\beq
\Gamma = \sum_n L^\dagger_n L_n\,.
\eeq
Indeed, as we will show, the damping matrix that we will determine
by means of \Eq{GammaVfinal}, can be written in the form
\beq
\label{GammaL}
\Gamma = L^{(W)\dagger}_e L^{(W)}_e +
\sum_{f = e,n,p} L^{(Z)\dagger}_f L^{(Z)}_f\,,
\eeq
with well-defined expressions for the $L$ matrices in terms
of integrals over the background particles distribution functions
that we will obtain from the self-energy calculation.

\subsection{Notation and conventions}
\label{sec:notation}

For definiteness we state precisely the notation and conventions
we use throughout.
The neutral-current couplings of the interaction Lagrangian that are
relevant to our calculation are given by
\beq
\label{LZ}
L_{Z} = -g_Z Z^\mu\left[\sum_a \overline\nu_{La}\gamma_\mu\nu_{La} + 
\overline e\gamma_\mu(a_e + b_e\gamma_5) e + J^{(Z)}_\mu\right]\,,
\eeq
where, in the standard model,
\begin{equation}
g_Z = g/(2\cos\theta_W)
\end{equation}
and
\beqa
\label{aebe}
a_e & = & -\frac{1}{2} + 2\sin^2\theta_W \,,\nonumber\\
b_e & = & \frac{1}{2}\,.
\eeqa
On the other hand, $J^{(Z)}_\mu$ is the nucleon neutral current,
which in terms of the quark fields
\beq
q = \left(\begin{array}{l}
u\\
d
\end{array}
\right)\,,
\eeq
is given by
\beq
J^{(Z)}_\mu = \overline q\gamma_\mu \frac{\tau_3}{2} q
- \overline q\gamma_\mu\gamma_5 \frac{\tau_3}{2}q 
- 2\sin^2\theta_W J^{(em)}_\mu\,,
\eeq
where $J^{(em)}_\mu$ is the electromagnetic current
\beq
J^{(em)}_\mu = \overline q\gamma_\mu \frac{\tau_3}{2} q
+ \frac{1}{6}\overline q\gamma_\mu q\,,
\eeq
and $\tau_{1,2,3}$ stand for the Pauli matrices.

We introduce the nucleon ($f = n,p$) neutral-current vertex function
$j^{(Z)}_{f\mu}(q)$, which is defined such that the matrix element of the
neutral-current between nucleon states is given by
\beq
\label{jzdef}
\langle f(p^\prime)|J^{(Z)}_\mu(0)|f(p)\rangle
= \overline u(p^\prime)j^{(Z)}_{f\mu}(p - p^\prime)u(p)\,.
\eeq
We parametrize $j^{(Z)}_{f\mu}(q)$ in the form
\beq
\label{jznucl}
j^{(Z)}_{f\mu}(q) = a_f\gamma_\mu + b_f\gamma_\mu\gamma_5
- i\frac{c_f}{2m_N}\sigma_{\mu\nu}q^\nu\,.
\eeq
In principle the parameters $a_f,b_f,c_f$ are $q^2$-dependent form factors.
For our purposes we will assume that it is valid to adopt their $q^2 = 0$
limiting value. In this case,
\beqa
\label{abcnucleons}
a_f & = & I_{3f} - 2\sin^2\theta_W Q_f\,,\nonumber\\
b_f & = & -I_{3f}g_A\,,\nonumber\\
c_f & = & I_{3f}[F^{(em)}_{2p}(0) - F^{(em)}_{2n}(0)] - 
2\sin^2\theta_W F^{(em)}_{2f}(0)\,,
\eeqa
where $Q_p = 1, Q_n = 0$, $I_{3p} = -I_{3n} = 1/2$ and
\beqa
F_{2p}^{(em)}(0) & = & 1.79\,,\nonumber\\
F_{2n}^{(em)}(0) & = & -1.71\,.
\eeqa
In addition we will discard the $c_f$ term since it contains a factor
of $q/m_N$ which gives a small contribution relative to the other terms.
For the charged current,
\beq
L_W  = -\left(\frac{g}{\sqrt{2}}\right)W^\mu \nu_L\gamma_\mu e_L + h.c.
\eeq
\section{Calculation of $\Gamma$ and the jump operators}
\label{sec:Gamma}

\subsection{Calculation of $\Sigma_{12}$}
\label{subsec:sigma12}

We consider first the contribution to $\Sigma_{12}(k)$
from diagram (b) in \Fig{fig:sm}. In the \emph{heavy} $Z$ limit,
only the diagonal elements of the $Z$ propagator are non-zero,
$\Delta^{(Z)}_{22\mu\nu}= -\Delta^{(Z)}_{11\mu\nu}= -g_{\mu\nu}/m^2_Z$,
and therefore only the terms with $A = 1, B = 2$ contribute.
Each fermion in the background contributes a term that we write in the form
\beqa
\label{SigmaZ1}
-i\left(\Sigma^{(Z,f)}_{12}(k)\right)_{ba} & = & -2K^{(Z)}_{ba}
\int\frac{d^4p^\prime}{(2\pi)^4}\frac{d^4p}{(2\pi)^4}
\gamma^\mu L iS^{(\nu_{La})}_{12}(k^\prime)\gamma^\nu L\nonumber\\
&&\mbox{}\times
\Tr\left(\gamma_\mu(a_f + b_f\gamma_5)iS^{(f)}_{12}(p^\prime)
\gamma_\nu(a_f + b_f\gamma_5)iS^{(f)}_{21}(p)
\right)\,,
\eeqa
where
\beq
K^{(Z)}_{ba} = \left(\frac{g^4_Z}{2m^4_Z}\right)\delta_{ab} =
\left(\frac{g^4}{32m^4_W}\right)\delta_{ab}\,,
\eeq
and
\beq
\label{kprime}
k^\prime \equiv p - p^\prime + k\,.
\eeq
The corresponding expression for the contribution from diagram (a)
can be obtained from \Eq{SigmaZ1} by making simple substitutions. Thus,
\beqa
\label{SigmaW1}
-i\left(\Sigma^{(W)}_{12}(k)\right)_{ba} & = & -2K^{(W)}_{ba}
\int\frac{d^4p^\prime}{(2\pi)^4}\frac{d^4p}{(2\pi)^4}
\gamma^\mu L iS^{(e)}_{12}(p^\prime)\gamma^\nu L\nonumber\\
&&\mbox{}\times
\Tr\left(\gamma_\mu L iS^{(\nu_{Le})}_{12}(k^\prime)
\gamma_\nu L iS^{(e)}_{21}(p)
\right)\,,
\eeqa
where
\beq
K^{(W)}_{ba} = \left(\frac{g^4}{8m^4_W}\right)\delta_{ae}\delta_{be}\,,
\eeq
which can in turn be rewritten in the form (the proof is given in
\Appendix{sec:identityBCD})
\beqa
\label{SigmaW2}
-i\left(\Sigma^{(W)}_{12}(k)\right)_{ba} & = & -2K^{(W)}_{ba}
\int\frac{d^4p^\prime}{(2\pi)^4}\frac{d^4p}{(2\pi)^4}
\gamma^\mu L iS^{(\nu_{Le})}_{12}(k^\prime)\gamma^\nu L\nonumber\\
&&\mbox{}\times
\Tr\left(\gamma_\mu L iS^{(e)}_{12}(p^\prime)
\gamma_\nu L iS^{(e)}_{21}(p)
\right)\,.
\eeqa
Therefore, in what follows we concentrate on the evaluation
of $\Sigma^{(Z,f)}_{12}(k)$ using \Eq{SigmaZ1}. The results
for $\Sigma^{(W)}_{12}(k)$ are obtained by making the replacements
\beqa
\label{replacements}
K^{(Z)}_{ba} & \rightarrow & K^{(W)}_{ba}\,,\nonumber\\
a_e = -b_e & \rightarrow & \frac{1}{2}\,,
\eeqa
in the results for $\Sigma^{(Z,e)}_{12}(k)$.

For the propagators of the internal fermion and neutrino lines
we adopt the same formulas used in \xRef{nsnuphidecoherence}.
Specifically, we express the components of the $f$ propagator matrices
in the form
\beqa
\label{fpropagators}
S^{(f)}_{21}(p) & = & -2\pi i\delta(p^2 - m^2_f)\sigma^{(f)}(p)
e^{x_f} n_F(x_f)\epsilon(p\cdot u)\,,\nonumber\\
S^{(f)}_{12}(p^\prime) & = & 2\pi i\delta(p^{\prime\,2} - m^2_f)
\sigma^{(f)}(p^\prime)n_F(x^\prime_f)\epsilon(p^\prime\cdot u)\,,
\eeqa
where
\beq
\label{sigmaf}
\sigma^{(f)}(q) = \lslash{q} + m_f\,,
\eeq
$n_F(z)$ is the fermion distribution function 
defined in \Eq{nzfermion}, $\epsilon(z) = \theta(z) - \theta(-z)$
where $\theta(z)$ is the step function, and we have defined
\beqa
x_f & = & \beta p\cdot u - \alpha_f\,,\nonumber\\
x^\prime_f & = & \beta p^\prime\cdot u - \alpha_f\,.
\eeqa
For the neutrino propagator, we neglect the effect of the non-zero neutrino
masses and/or dispersion relations in the calculation of $\Sigma_{12}$
as in \xRef{nsnuphidecoherence}. In this case the neutrino propagator
is diagonal in flavor space, with all the elements actually being the
same since all the neutrinos have the same mass (zero) and the same
chemical potential. Specifically,
\beq
\label{nupropagator}
(S^{(\nu_{La})}_{12}(k^\prime)) = 2\pi i\delta(k^{\prime\,2})
\sigma^{(\nu)}(k^\prime)n_F(x^\prime_\nu)\epsilon(k^\prime\cdot u)\,,
\eeq
where
\beq
\label{xnu}
x^\prime_\nu = \beta k^\prime\cdot u - \alpha_\nu\,,
\eeq
and
\beq
\label{sigmanu}
\sigma^{(\nu)}(k^\prime) = L\lslash{k}^\prime\,.
\eeq
The distribution function for the fermion $f$ and neutrino are denoted by
$f_{f}$ and $f_\nu$, respectively, with
\beq
\label{fx}
f_f(\epsilon) = \frac{1}{e^{\beta\epsilon - \alpha_f} + 1}\,,
\eeq
and an analogous formula for $f_\nu$,
while corresponding formulas for the antiparticles, $f_{\bar f,\bar\nu}$,
are given by reversing the sign of $\alpha_{f,\nu}$.

We will denote by $\Sigma^{(Z,f)}_i$ the contribution
to $\Sigma_{i}$ due to the $\Sigma^{(Z,f)}_{12}$ term we are considering.
That is, from \Eq{Sigmaidef1},
\beq
\label{SigmaideffZ}
\Sigma^{(Z,f)}_i(k) = \frac{\Sigma^{(Z,f)}_{12}}{2in_F(x_\nu)}\,.
\eeq
In the following steps we mimic the procedure we used in
\xRef{nsnuphidecoherence}, and therefore we omit here some of the details.
Thus, we let $k^\prime$ be an arbitrary four-momentum variable in the
integral expression for $\Sigma^{(Z,f)}_{12}$ but insert the factor
$\delta^{(4)}(k^\prime + p^\prime - p - k)$ and integrating over $k^\prime$.
Then carrying out the integral over $k^{\prime\,0}$, with the help
of the delta function,
\beqa
\label{SigmaiZfba2}
\left(\Sigma^{(Z,f)}_i(k)\right)_{ba} & = & -K^{(Z)}_{ba}
\int\frac{d^4p^\prime}{(2\pi)^3}
\frac{d^4p}{(2\pi)^3}
\frac{d^3\kappa^\prime}{(2\pi)^3 2\omega_{\kappa^\prime}}
\delta(p^{\prime\,2} - m^2_f)\delta(p^2 - m^2_f)
\epsilon(p\cdot u)\epsilon(p^\prime\cdot u)
\nonumber\\
&& \times (2\pi)^4\left\{
\delta^{(4)}(k + p - k^\prime - p^\prime)
N_{\mu\nu}(p,p^\prime)M^{\mu\nu}(k^\prime)E_\nu\right.
\nonumber\\
&& - \left.\delta^{(4)}(k + p + k^\prime - p^\prime)
N_{\mu\nu}(p,p^\prime)M^{\mu\nu}(-k^\prime)E_{\bar\nu}
\right\}\,,
\eeqa
where
\beqa
\label{MN}
M^{\mu\nu}(k^\prime) & = & \gamma^\mu L\sigma^{(\nu)}(k^\prime)\gamma^\nu L\,,
\nonumber\\
N_{\mu\nu}(p,p^\prime) & = &
\Tr\left(\gamma_\mu(a_f + b_f\gamma_5)\sigma^{(f)}(p^\prime)
\gamma_\nu(a_f + b_f\gamma_5)\sigma^{(f)}(p)\right)\,,
\eeqa
and
\beqa
\label{Enu}
E_\nu & = & n_F(x_f)(1 - n_F(x^\prime_f)) -
f_\nu(\omega_{\kappa^\prime})(n_F(x_f) - n_F(x^\prime_f))
\,,\nonumber\\
E_{\bar\nu} & = & n_F(x^\prime_f)(1 - n_F(x_f)) +
f_{\bar\nu}(\omega_{\kappa^\prime})(n_F(x_f) - n_F(x^\prime_f))\,,
\eeqa
with
\beqa
k^{\prime\,\mu} & = & (\omega_{\kappa^\prime},\vec\kappa^\prime)
\,,\nonumber\\
\omega_{\kappa^\prime} & = & |\vec\kappa^\prime|\,.
\eeqa
Next carrying out the integration over $p^0, p^{\prime\,0}$ in a similar way,
we obtain
\beqa
\label{SigmaiZfba3}
\left(\Sigma^{(Z,f)}_i(k)\right)_{ba} & = & 
-K^{(Z)}_{ba}\int\frac{d^3p^\prime}{(2\pi)^3 2E_{p^\prime}}
\frac{d^3p}{(2\pi)^3 2E_{p}}
\frac{d^3\kappa^\prime}{(2\pi)^3 2\omega_{\kappa^\prime}}
\nonumber\\
&& \times (2\pi)^4\left\{
\delta^{(4)}(k + p - k^\prime - p^\prime)
N_{\mu\nu}(p,p^\prime)M^{\mu\nu}(k^\prime)E_{\nu,++}
\right.\nonumber\\
&& - \left.\delta^{(4)}(k - p - k^\prime - p^\prime)
N_{\mu\nu}(-p,p^\prime)M^{\mu\nu}(k^\prime)E_{\nu,-+}
\right.\nonumber\\
&& - \left.\delta^{(4)}(k + p + p^\prime - k^\prime)
N_{\mu\nu}(p,-p^\prime)M^{\mu\nu}(k^\prime)E_{\nu,+-}
\right.\nonumber\\
&& + \left.\delta^{(4)}(k + p^\prime - k^\prime - p)
N_{\mu\nu}(-p,-p^\prime)M^{\mu\nu}(k^\prime)E_{\nu,--}
\right.\nonumber\\
&& - \left.\delta^{(4)}(k + p + k^\prime - p^\prime)
N_{\mu\nu}(p,p^\prime)M^{\mu\nu}(-k^\prime)E_{\bar\nu,++}
\right.\nonumber\\
&& + \left.\delta^{(4)}(k + k^\prime - p^\prime - p)
N_{\mu\nu}(-p,p^\prime)M^{\mu\nu}(-k^\prime)E_{\bar\nu,-+}
\right.\nonumber\\
&& + \left.\delta^{(4)}(k + p + p^\prime + k^\prime)
N_{\mu\nu}(p,-p^\prime)M^{\mu\nu}(-k^\prime)E_{\bar\nu,+-}
\right.\nonumber\\
&& - \left.\delta^{(4)}(k + p^\prime + k^\prime - p)
N_{\mu\nu}(-p,-p^\prime)M^{\mu\nu}(-k^\prime)E_{\bar\nu,--}
\right\}\,,
\eeqa
with
\beq
p^\mu = (E_p,\vec p)\,,\qquad
E_p = \sqrt{{\vec p}^{\;2} + m^2_f}\,,
\eeq
and similarly for $p^{\prime\,\mu}$. In \Eq{SigmaiZfba3} we have introduced
the factors $E_{\nu,\lambda\lambda^\prime}$ and
$E_{\bar\nu,\lambda\lambda^\prime}$
(with $\lambda,\lambda^\prime$ being $\pm$), which are defined as follows,
\beqa
E_{\nu,\lambda\lambda^\prime} = \left.E_\nu\right|_{
p^0 = \lambda E_p,\,p^{\prime\,0} = \lambda^\prime E_{p^\prime}}\,,
\eeqa
and similarly for $E_{\bar\nu,\lambda\lambda^\prime}$.
The explicit formulas are given in \Table{table:processes}.
\begin{table}
\begin{center}
  \begin{tabular}{|l|l|}
    \hline
  $E_{\nu,++} = \fEp(1 - \fEpp) - \fnu(\fEp - \fEpp)$ &
  $\nu_a(k) + f(p) \leftrightarrow \nu_a(k^\prime) + f(p^\prime)$\\
  $E_{\nu,-+} = (1 - \fbEp)(1 - \fEpp) - \fnu(1 - \fbEp - \fEpp)$ &
  $\nu_a(k) \leftrightarrow \nu_a(k^\prime) + \bar f(p) + f(p^\prime)$\\
  $E_{\nu,+-} = \fEp\fbEpp - \fnu(\fEp + \fbEpp - 1)$ &
  $\nu_a(k) + f(p) + \bar f(p^\prime) \leftrightarrow \nu_a(k^\prime)$\\
  $E_{\nu,--} = (1 - \fbEp)\fbEpp - \fnu(\fbEpp - \fbEp)$ &    
  $\nu_a(k) + \bar f(p^\prime) \leftrightarrow
  \nu_a(k^\prime) + \bar f(p)$\\
  $E_{\bar\nu,++} = (1 - \fEp)\fEpp + \fnub(\fEp - \fEpp)$ &
  $\nu_a(k) + \bar\nu_a(\bar k^\prime) + f(p)\leftrightarrow f(p^\prime)$\\
  $E_{\bar\nu,-+} = \fbEp\fEpp + \fnub(1 - \fbEp - \fEpp)$ &
  $\nu_a(k) + \bar\nu_a(\bar k^\prime) \leftrightarrow
  \bar f(p) + f(p^\prime)$\\
  $E_{\bar\nu,+-} = (1 - \fEp)(1 - \fbEpp) + \fnub(\fEp + \fbEpp - 1)$ &
  $\nu_a(k) + \bar \nu_a(\bar k^\prime) +
  f(p) + \bar f(p^\prime) \leftrightarrow 0$\\
  $E_{\bar\nu,--} = \fbEp(1 - \fbEpp) + \fnub(\fbEpp - \fbEp)$ &
  $\nu_a(k) + \bar\nu_a(\bar k^\prime) + \bar f(p^\prime) \leftrightarrow
  \bar f(p)$\\
  \hline
\end{tabular}
\caption{Correspondence between the $E_{\nu,\lambda\lambda^\prime}$
  and $E_{\bar\nu,\lambda\lambda^\prime}$ factors defined in \Eq{Enu},
  and the process that contributes to the $\nu(k)$ damping via
  \Eq{SigmaiZfba2}. To simplify the notation we are using
    the shorthands shown in \Eq{fshorthand} for the various distribution
    functions.
  \label{table:processes}
}
\end{center}
\end{table}
To simplify the notation in the formulas summarized in \Table{table:processes}
we have introduce the shorthands
\beqa
\label{fshorthand}
f = f_f(E_p), \quad f^\prime = f_f(E_{p^\prime}), \quad
f^\prime_\nu = {f_\nu(\omega_{\kappa^\prime})}\nonumber\\
\bar f = f_{\bar f}(E_p), \quad \bar f^\prime = f_{\bar f}(E_{p^\prime}), \quad
\bar f^\prime_\nu = f_{\bar\nu}(\omega_{\kappa^\prime})\,.
\eeqa
The formulas for $E_{\bar\nu,\lambda\lambda^\prime}$ are obtained from
those for $E_{\nu,\lambda\lambda^\prime}$ by making the replacement
$\fnu \rightarrow (1 - \fnub)$.

As discussed in \xRef{nsnuphidecoherence}, each of the terms that appear
within the bracket in \Eq{SigmaiZfba3} corresponds to a particular
non-forward scattering process, and its inverse, for example
\beq
\label{process1}
\nu_{a}(k) + f(p) \leftrightarrow \nu_a(k^\prime) + f(p^\prime)\,,
\eeq
as well as the processes obtained by crossing
$f(p), f(p^\prime),\nu_a(k^\prime)$. For $\omega > 0$,
the only processes that are kinematically accessible are the one shown above,
and the following one,
\beq
\nu_a(k) + \bar f(p^\prime) \rightarrow \nu_a(k^\prime) + \bar f(p)\,.
\eeq
These correspond to the the first and the fourth
terms, respectively, in the list of terms that appear within the
bracket in \Eq{SigmaiZfba3}. Alternatively, for $\omega < 0$, the only
kinematically accessible processes are
\beqa
\bar\nu_{a}(k) + f(p^\prime) & \rightarrow &
\bar \nu_a(k^\prime) + f(p)\,,\nonumber\\
\bar\nu_{a}(k) + \bar f(p) & \rightarrow &
\bar\nu_a(k^\prime) + \bar f(p^\prime)\,,
\eeqa
which correspond to the fifth and eighth terms within the
bracket in \Eq{SigmaiZfba3}. In addition we will assume that there
are no neutrinos or antineutrinos in the background, therefore we set
$f_\nu$ and $f_{\bar\nu}$ to zero. Then,
\beqa
\label{SigmaiZfba4}
\left(\Sigma^{(Z,f)}_i(k)\right)_{ba} & = & 
-K^{(Z)}_{ba}\int\frac{d^3p^\prime}{(2\pi)^3 2E_{p^\prime}}
\frac{d^3p}{(2\pi)^3 2E_{p}}
\frac{d^3\kappa^\prime}{(2\pi)^3 2\omega_{\kappa^\prime}}
\nonumber\\
&& \times  (2\pi)^4\left\{
\delta^{(4)}(k + p - k^\prime - p^\prime)
N_{\mu\nu}(p,p^\prime)M^{\mu\nu}(k^\prime)
\left[f_{f}(E_p)(1 - f_{f}(E_{p^\prime}))\right]\right.\nonumber\\
&& + \left.\delta^{(4)}(k + p^\prime - k^\prime - p)
N_{\mu\nu}(-p,-p^\prime)M^{\mu\nu}(k^\prime)
\left[(1 - f_{\bar f}(E_p))f_{\bar f}(E_{p^\prime})\right]
\right\}\,.\nonumber\\
&& + \left.\delta^{(4)}(k + p + k^\prime - p^\prime)
N_{\mu\nu}(p,p^\prime)M^{\mu\nu}(k^\prime)
\left[(1 - f_{f}(E_{p}))f_{f}(E_{p^\prime})\right]\right.\nonumber\\
&& + \left.\delta^{(4)}(k + p^\prime + k^\prime - p)
N_{\mu\nu}(-p,-p^\prime)M^{\mu\nu}(k^\prime)
\left[f_{\bar f}(E_{p})(1 - f_{\bar f}(E_{p^\prime}))\right]
\right\}\,,\nonumber\\
\eeqa
where we have used the fact that
$M^{\mu\nu}(-k^\prime) = -M^{\mu\nu}(k^\prime)$ and, as we have mentioned,
if $\omega > 0$ only the first two terms in the bracket contribute,
while for $\omega < 0$ only the last two contribute.

\subsection{Calculation of $n\cdot V_i$}

As already stated in \Section{sec:preliminaries},
the contribution to $\left(V^\alpha_i(\omega,\vec\kappa)\right)_{ba}$,
which we denote by $\left(V^{(Z,f)\alpha}_i(\omega,\vec\kappa)\right)_{ba}$,
is obtained by substituting \Eq{SigmaiZfba4} in \Eq{ViZfSigmailoopformula}.
It then follows that the formula for
$\left(V^{(Z,f)\alpha}_i(\omega,\vec\kappa)\right)_{ba}$
is obtained from \Eq{SigmaiZfba4} by making the replacement
\beq
M^{\mu\nu}(k^\prime) \rightarrow L^{\alpha\mu\nu}(k^\prime)\,.
\eeq
where
\beqa
\label{L}
L^{\alpha\mu\nu}(k^\prime) & \equiv &
\frac{1}{2}\Tr\gamma^\alpha M^{\mu\nu}(k^\prime)\nonumber\\
& = & \frac{1}{2}\Tr\gamma^\alpha \gamma^\mu L\sigma^{(\nu)}(k^\prime)
\gamma^\nu L\nonumber\\
& = & \frac{1}{2}\Tr L\gamma^\alpha \gamma^\mu \lslash{k}^\prime\gamma^\nu\,.
\eeqa
That is,
\beqa
\label{ViZfba1}
\left(V^{(Z,f)\alpha}_i(\omega,\vec\kappa)\right)_{ba} & = & 
-K^{(Z)}_{ba}\int\frac{d^3p^\prime}{(2\pi)^3 2E_{p^\prime}}
\frac{d^3p}{(2\pi)^3 2E_{p}}
\frac{d^3\kappa^\prime}{(2\pi)^3 2\omega_{\kappa^\prime}}
\nonumber\\
&& \times  (2\pi)^4\left\{
\delta^{(4)}(k + p - k^\prime - p^\prime)
N_{\mu\nu}(p,p^\prime)L^{\alpha\mu\nu}(k^\prime)
\left[f_{f}(E_p)(1 - f_{f}(E_{p^\prime}))\right]\right.\nonumber\\
&& + \left.\delta^{(4)}(k + p^\prime - k^\prime - p)
N_{\mu\nu}(-p,-p^\prime)L^{\alpha\mu\nu}(k^\prime)
\left[(1 - f_{\bar f}(E_p))f_{\bar f}(E_{p^\prime})\right]\right.\nonumber\\
&& + \left.\delta^{(4)}(k + p + k^\prime - p^\prime)
N_{\mu\nu}(p,p^\prime)L^{\alpha\mu\nu}(k^\prime)
\left[(1 - f_{f}(E_{p}))f_{f}(E_{p^\prime})\right]\right.\nonumber\\
&& + \left.\delta^{(4)}(k + p^\prime + k^\prime - p)
N_{\mu\nu}(-p,-p^\prime)L^{\alpha\mu\nu}(k^\prime)
\left[f_{\bar f}(E_{p})(1 - f_{\bar f}(E_{p^\prime}))\right]
\right\}\,.
\eeqa
The traces involved in \Eqs{MN}{L} are easily
evaluated by means of the standard formulas. After
some straightforward algebra, this procedure leads to
\beqa
\label{ViZfba2}
\left(V^{(Z,f)\alpha}_i(\omega,\vec\kappa)\right)_{ba} & = & 
-8K^{(Z)}_{ba}\int\frac{d^3p^\prime}{(2\pi)^3 2E_{p^\prime}}
\frac{d^3p}{(2\pi)^3 2E_{p}}
\frac{d^3\kappa^\prime}{(2\pi)^3 2\omega_{\kappa^\prime}}
v^\alpha(k^\prime,p,p^\prime)
\nonumber\\
&& \times  (2\pi)^4\left\{
\delta^{(4)}(k + p - k^\prime - p^\prime)
\left[f_{f}(E_p)(1 - f_{f}(E_{p^\prime}))\right]\right.\nonumber\\
&& + \left.\delta^{(4)}(k + p^\prime - k^\prime - p)
\left[(1 - f_{\bar f}(E_p))f_{\bar f}(E_{p^\prime})\right]
\right\}\,.\nonumber\\
&& + \left.\delta^{(4)}(k + p + k^\prime - p^\prime)
\left[(1 - f_{f}(E_{p}))f_{f}(E_{p^\prime})\right]\right.\nonumber\\
&& + \left.\delta^{(4)}(k + p^\prime + k^\prime - p)
\left[f_{\bar f}(E_{p})(1 - f_{\bar f}(E_{p^\prime}))\right]
\right\}\,,
\eeqa
where
\beq
v^\alpha(k^\prime,p,p^\prime) \equiv -m^2_f(a^2_f - b^2_f)k^{\prime\,\alpha} +
(a_f - b_f)^2(k^\prime\cdot p^\prime)p^\alpha +
(a_f + b_f)^2(k^\prime\cdot p)p^{\prime\,\alpha}\,.
\eeq
The quantities that enter in the formula for $\Gamma$ are then,
\beqa
\label{VZffinal1}
n\cdot\left(V^{(Z,f)}_i(\kappa,\vec\kappa)\right)_{ba} & = &
-4K^{(Z)}_{ba}
\left\{-m^2_f(a^2_f - b^2_f)(I^{(f)}_0 + I^{(\bar f)}_0)\right.
\nonumber\\
&&\mbox{} + \left.(a_f + b_f)^2(I^{(f)}_1 + I^{(\bar f)}_2)
\right.
\nonumber\\
&&\mbox{} + \left.(a_f - b_f)^2(I^{(f)}_2 + I^{(\bar f)}_1)
\right\}\,,
\eeqa
\beqa
\label{VZffinal2}
n\cdot\left(V^{(Z,f)}_i(-\kappa,-\vec\kappa)\right)_{ba} & = &
-4K^{(Z)}_{ba}
\left\{-m^2_f(a^2_f - b^2_f)(I^{(f)}_0 + I^{(\bar f)}_0)\right.
\nonumber\\
&&\mbox{} + \left.(a_f + b_f)^2(I^{(f)}_2 + I^{(\bar f)}_1)
\right.
\nonumber\\
&&\mbox{} + \left.(a_f - b_f)^2(I^{(f)}_1 + I^{(\bar f)}_2)
\right\}\,,
\eeqa
where we have introduced the integrals $I^{(f,\bar f)}_0$, which
for either case ($x = f,\bar f$) is defined as
\beqa
\label{I0xfinal}
I^{(x)}_{0} & = & \frac{2}{\omega_\kappa}
\int\frac{d^3\kappa^\prime}{(2\pi)^3 2\omega_{\kappa^\prime}}
\frac{d^3p^\prime}{(2\pi)^3 2E_{p^\prime}}
\frac{d^3p}{(2\pi)^3 2E_{p}}\nonumber\\
&& \times (2\pi)^4\delta^{(4)}(p + q - p^\prime)
f_{x}(E_{p})\left(1 - f_{x}(E_{p^\prime})\right)k\cdot k^\prime\,,
\eeqa
while $I^{(f,\bar f)}_{1,2}$ are the same integrals defined in
\xRef{nsnuphidecoherence}, which we reproduce here for convenience,
\beqa
\label{Ix12final}
I^{(x)}_1 & = & \frac{2}{\omega_\kappa}\int\frac{d^3p}{(2\pi)^3 2E_{p}}
\frac{d^3p^\prime}{(2\pi)^3 2E_{p^\prime}}
\frac{d^3\kappa^\prime}{(2\pi)^3 2\omega_{\kappa^\prime}}
\nonumber\\
&& \times (2\pi)^4
\delta^{(4)}(p + k - p^\prime - k^\prime)
f_{x}(E_{p})\left(1 - f_{x}(E_{p^\prime})\right)
(p\cdot k^\prime)^2\,,\nonumber\\
I^{(x)}_2 & = & \frac{2}{\omega_\kappa}\int\frac{d^3p}{(2\pi)^3 2E_{p}}
\frac{d^3p^\prime}{(2\pi)^3 2E_{p^\prime}}
\frac{d^3\kappa^\prime}{(2\pi)^3 2\omega_{\kappa^\prime}}
\nonumber\\
&& \times (2\pi)^4
\delta^{(4)}(p + k - p^\prime - k^\prime)
f_{x}(E_{p})\left(1 - f_{x}(E_{p^\prime})\right)(p\cdot k)^2\,.
\eeqa
In these integral formulas, we understand that $k$ is set to
\beq
k^\mu = (\omega_\kappa,\vec\kappa)\,,
\eeq
with $\omega_\kappa = \kappa$, and similarly for $k^\prime$.

\subsection{Formula for $\Gamma$}

We can now obtain the explicit formula for the damping matrix in terms of the
integrals $I^{(x)}_{0,1,2}$.
The damping matrix is given by \Eq{GammaVfinal},
where $n\cdot V^{(Z,f)}_i(\kappa,\vec\kappa)$ and
$n\cdot V^{(Z,f)}_i(-\kappa,-\vec\kappa)$ are given above
in \Eqs{VZffinal1}{VZffinal2} while the formulas for
$n\cdot V^{(W)}(\kappa,\vec\kappa)$ and $n\cdot V^{(W)}(-\kappa,-\vec\kappa)$
are obtained from the corresponding formulas for $n\cdot V^{(Z,f)}_i$
by making the replacement indicated in \Eq{replacements}.
Therefore, for the neutrinos,
\beq
\label{Gammanugamma}
\Gamma^{(\nu)}_{ba} = \left(\frac{g^2}{2m^2_W}\right)^2\left[\gamma^{(W)}_e
  \delta_{be}\delta_{ae}
  + \left(\sum_f \gamma^{(Z)}_f\right)\delta_{ba}\right]\,,
\eeq
where
\beqa
\label{gamma}
\gamma^{(Z)}_f & = &
-m^2_f(a^2_f - b^2_f)(I^{(f)}_0 + I^{(\bar f)}_0)
+ (a_f + b_f)^2(I^{(f)}_1 + I^{(\bar f)}_2)
+ (a_f - b_f)^2(I^{(f)}_2 + I^{(\bar f)}_1)\,,\nonumber\\
\gamma^{(W)}_e & = & (I^{(e)}_2 + I^{(\bar e)}_1)\,.
\eeqa
For the antineutrinos, the formula for $\Gamma^{(\bar\nu)}$ is similar
to \Eq{Gammanugamma}, with $\gamma^{(W,Z)}_f \rightarrow \bar\gamma^{(W,Z)}_f$,
where
\beqa
\label{gammabar}
\bar\gamma^{(Z)}_f & = &
-m^2_f(a^2_f - b^2_f)(I^{(f)}_0 + I^{(\bar f)}_0)
+ (a_f + b_f)^2(I^{(f)}_2 + I^{(\bar f)}_1)
+ (a_f - b_f)^2(I^{(f)}_1 + I^{(\bar f)}_2)\,,\nonumber\\
\bar\gamma^{(W)}_e & = & (I^{(e)}_1 + I^{(\bar e)}_2)\,.
\eeqa

\subsection{Formula for the jump operators and decoherence terms}
\label{subsec:decoherence}

As already explained in \Section{sec:decoherence}, the proposal for identifying
the jump operators is based on writing $\Gamma$ as sum
of terms of the form $L^\dagger L$.
Looking at \Eq{GammaVfinal} we see that we can write $\Gamma$ in the form
given in \Eq{GammaL},
%
%
with
\beqa
\label{Lgamma}
(L^{(Z)}_f)_{ba} & = & \left(\frac{g^2}{2m^2_W}\right)\sqrt{\gamma^{(Z)}_f}
\delta_{ba}\,,\nonumber\\
(L^{(W)}_e)_{ba} & = & \left(\frac{g^2}{2m^2_W}\right)\sqrt{\gamma^{(W)}_e}
\delta_{be}\delta_{ae}\,,
\eeqa
or in matrix notation,
\beqa
\label{Lgammamatrix}
L^{(Z)}_f & = & \left(\frac{g^2}{2m^2_W}\right)\sqrt{\gamma^{(Z)}_f} I
\,,\nonumber\\
L^{(W)}_e & = & \left(\frac{g^2}{2m^2_W}\right)\sqrt{\gamma^{(W)}_e} I_e\,,
\eeqa
where $I$ is the identity matrix and
\beq
\label{Ie}
I_e = \left(\begin{array}{ccc}
  1 & 0 & 0\\
  0 & 0 & 0\\
  0 & 0 & 0
\end{array}\right)
\eeq
For the antineutrinos, the result is similar, with
$\gamma^{(W,Z)}_f \rightarrow \bar\gamma^{(W,Z)}_f$.
\Eq{Lgamma}, together with \Eqs{gamma}{gammabar}
are the central results of the present work.
We then assert that the damping effects
of the non-forward scattering processes are properly taken
into account in the context of
the evolution equation for the flavor density matrix,
\beq
\label{eveqL}
\partial_t\rho = -i[H_r,\rho] + \sum_{\substack{X = Z,W\\f = e,n,p}}
\left\{L^{(X)}_f \rho L^{(X)\dagger}_f -
\frac{1}{2}L^{(X)\dagger}_f L^{(X)}_f\rho -
\frac{1}{2}\rho L^{(X)\dagger}_f L^{(X)}_f\right\}\,,
\eeq
with $L^{(W,Z)}_f$ given in \Eq{Lgammamatrix}.
For reference, we refer to the terms involving the jump operators
on the right-hand side of \Eq{eveqL} as the decoherence terms.

Since the $L^{(Z)}_f$ terms are proportional to the identity matrix,
they all drop out of \Eq{eveqL}. The evolution equation
reduces to
\beq
\label{eveqLW}
\partial_t\rho = -i[H_r,\rho] + D\,,
\eeq
where
\beq
\label{D}
D = 2\gamma\left\{I_e \rho I_e - \frac{1}{2}I_e \rho -
\frac{1}{2}\rho I_e\right\}\,,
\eeq
with
\beq
\gamma = \frac{1}{2}\left(\frac{g^2}{2m^2_W}\right)^2 \gamma^{(W)}_e\,.
\eeq
Thus, the decoherence terms are driven by $\gamma^{(W)}_e$ alone.
However, it should be kept in mind that this result holds if all the neutrinos
involved have the same neutral current couplings. In the presence of
non-universality (e.g., neutrino mixing involving non-active
neutrinos), the $L^{(Z)}_f$ terms are not proportional to the identity
matrix and \Eq{eveqLW} does not hold. In \Section{sec:evaluationintegrals}
we evaluate the integrals required to determine $\gamma^{(W)}_e$ in some
illustrative cases. Keeping the previous comment in mind, for completenesss
we include those for $\gamma^{(Z)}_f$ as well.
\section{Two-generation example}
\label{sec:twogenexample}

As an example application, for definiteness we consider the standard
two-generation case in a normal matter background.
The Wolfenstein term must be included in the Hamiltonian.
Our discussion resembles the one in previous works
that consider the decoherence effects, in which $\vec D$ is an unknown
and treated at a phenomenological level. In those
contexts the working assumption is that the decoherence terms are diagonal in
the basis of the effective mass eigenstates. This is not the case
with the $\vec D$ that we have obtained, and
the question we address here is how to take into account to calculate
the survival and transition probablities in the density matrix context.

We work in the flavor basis. The density matrix satisfies
\Eq{eveqLW}, with the initial normalization condition
\beq
\label{rhotzero}
\Tr \rho(0) = 1\,.
\eeq
Up to a term proportional to identity matrix that does not contribute
to the commutator, the Hamiltonian can be written in the form
\beq
H_{r} = \frac{1}{2}\vec \sigma\cdot \vec h\,,
\eeq
with
\beqa
\label{h}
\vec h & = & \left(\frac{\Delta m^2_{21}}{2\kappa}\sin 2\theta,\; 0,\;
-\frac{\Delta m^2_{21}}{2\kappa}\cos 2\theta + V_e\right)\,.
\eeqa
Here, $\Delta m^2_{21} = m^2_2 - m^2_1$ and 
$V_e$ is the Wolfenstein potential for electron neutrinos
$V_e = \sqrt{2}G_F n_e$, where $n_e$ is the total electron number density.
It is convenient to write
\beq
\label{hn}
\vec h = h\vec n\,,
\eeq
where $h$ is the magnitude of $\vec h$,
\beq
\label{hmag}
h = \frac{\Delta^2_m}{2\kappa}\,,
\eeq
with
\beq
\label{Deltam}
\Delta^2_m \equiv \sqrt{(\Delta m^2_{21}\sin 2\theta)^2 +
  (\Delta m^2_{21}\cos 2\theta - 2\kappa V_e)^2}\,,
\eeq
and $\vec n$ is the unit vector along $\vec h$. We also introduce
the vector with components
\beq
\vec e_3 = (0,0,1)\,,
\eeq
and define
\beqa
\label{cos2thetamdef}
\label{cos2thetam}
\label{sin2thetam}
\cos 2\theta_m & = & -\vec e_3\cdot\vec n =
\frac{1}{\Delta^2_m}
\left(\Delta m^2_{21}\cos 2\theta - 2\kappa V_e\right)\,,\nonumber\\
\sin^2 2\theta_m & = & 1 - (\vec e_3\cdot\vec n)^2 =
\left(\frac{\Delta m^2_{21}\sin 2\theta}{\Delta^2_m}\right)^2\,.
\eeqa

Parametrizing $\rho$ in the form
\beq
\label{rhoparametrization}
\rho = \frac{1}{2}\left(\sigma_0\rho_0 + \vec\sigma\cdot\vec\rho\right)\,,
\eeq
where $\sigma_0$ is the unit matrix and $\vec\sigma$ the Pauli matrices,
the evolution equation \Eq{eveqLW} gives
\beqa
\label{eveqourcase}
\partial_t\rho_0 & = & 0\,,\nonumber\\
\partial_t\vec\rho & = & h(\vec n\times\vec\rho) + \vec D\,,
\eeqa
where
\beq
\label{vecDourcase}
\vec D = -\frac{\gamma}{2}\vec\rho_\perp\,,
\eeq
with
\beq
\label{rhoperp}
\vec\rho_\perp = \vec\rho - (\vec e_3\cdot\vec\rho)\vec e_3\,.
\eeq
\Eq{eveqourcase} implies that $\rho_0$ is constant, and from
\Eq{rhotzero}
\beq
\rho_0(t) = 1\,.
\eeq

To solve the equation for $\vec\rho$, let us consider briefly
the equation with $\vec D = 0$,
\beq
\label{eveqD=0}
\partial_t\vec\rho = \vec h\times\vec \rho\,.
\eeq
Decomposing $\rho$ into its longitudinal and transverse components to
$\vec n$,
\beq
\vec\rho = \vec\rho_\ell + \vec\rho_t\,,
\eeq
where
\beqa
\vec\rho_\ell & \equiv & (\vec n\cdot\vec\rho)\vec n\,,\nonumber\\
\vec\rho_t & \equiv & \vec\rho - (\vec n\cdot\vec\rho)\vec n\,,
\eeqa
\Eq{eveqD=0} then implies that $\rho_\ell$ is constant,
\beq
\label{standardrholong}
\rho_\ell(t) = \rho_\ell(0)\,,
\eeq
while for $\vec\rho_t$,
\beq
\label{eveqtrans}
\partial_t \vec\rho_t = h\vec n\times\vec\rho_t\,.
\eeq
This is easily solved,
\beq
\label{rhotpure}
\vec\rho_t = \cos(ht)\vec\rho_t(0) + \sin(ht)\vec n\times\vec\rho_t(0)\,,
\eeq
so that
\beq
\label{rhosolD=0}
\vec\rho(t) = \vec\rho(0) + (\cos(ht) - 1)\vec\rho_t(0) +
\sin(ht)\vec n\times\vec\rho_t(0)\,,
\eeq

Going back to \Eq{vecDourcase}, the point is that $\vec\rho_\perp$ mixes
$\vec\rho_{\ell}$ and $\vec\rho_{t}$. In order to obtain a simple solution,
albeit approximate but nevertheless useful, we will treat
this mixing in a perturbative spirit. Thus we express $\vec\rho_\perp$
in the form
\beq
\label{rhoperpexpansion}
\vec\rho_\perp = a\vec\rho_\ell + b\vec\rho_t + OT\,,
\eeq
where $OT$ stands for other terms that we assume can be neglected as
a first approximation. Using \Eq{cos2thetamdef}, a simple calculation
then yields
\beqa
a & = & \sin^2 2\theta_m\,,\nonumber\\
b & = & \frac{1}{2}(1 + \cos^2 2\theta_m)\,.
\eeqa
Within this approximation, \Eqs{vecDourcase}{eveqourcase} give
\beq
\partial_t\vec\rho = h(\vec n\times\vec\rho) - \gamma_\ell\vec\rho_\ell -
\gamma_t\vec\rho_t\,,
\eeq
or
\beqa
\partial_t\rho_\ell & = & -\gamma_\ell\rho_\ell\,,\nonumber\\
\partial_t\vec\rho_t & = & h(\vec n\times\vec\rho_t) - \gamma_t\vec\rho_t\,,
\eeqa
where
\beqa
\label{gammaellt}
\gamma_\ell & = & n^2_1\gamma = \gamma\sin^2 2\theta_m\,,\nonumber\\
\gamma_t & = & \frac{1}{2}(1 + n^2_3)\gamma =
\frac{\gamma}{2}(1 + \cos^2 2\theta_m)\,.
\eeqa

For $\rho_\ell(0)$ we then have
\beq
\rho_\ell(t) = e^{-\gamma_\ell t}\rho_\ell(0)\,.
\eeq
A simple way to obtain the solution for $\vec\rho_t$ is to put
$\vec\rho_t = e^{-\gamma_t t}\vec\rho^{\,\prime}_t$\,,
so that the equation for $\vec\rho^{\,\prime}_t$ becomes the same
as the decoherence-free case. Thus we obtain
\beq
\vec\rho(t) = e^{-\gamma_\ell t}\vec\rho(0) +
\left[e^{-\gamma_t t}\cos(ht) - e^{-\gamma_\ell t}\right]
\vec\rho_t(0) + e^{-\gamma_t t}\sin(ht)\vec n\times\vec\rho_t(0)\,.
\eeq
Of course for $\gamma_{\ell,t} = 0$ we recover the decoherence-free
solution \Eq{rhosolD=0}.

As an example, suppose that initially
\beq
\rho(0) = \left(\begin{array}{cc}
1 & 0\\ 0 & 0
\end{array}\right) = \frac{1}{2}\left(1 + \vec\sigma\cdot\vec e_3\right)\,,
\eeq
which corresponds to $\vec\rho(0) = \vec e_3$. Then,
\beq
\label{vecrhosolexampledecoherence}
\vec\rho(t) = e^{-\gamma_\ell t}\vec e_3 +
\left[e^{-\gamma_t t}\cos(ht) - e^{-\gamma_\ell t}\right]
     [\vec e_3 - (\vec n\cdot\vec e_3)\vec n] +
     e^{-\gamma_t t}\sin(ht)\vec n\times\vec e_3\,.
\eeq
The survival and transition probabilities

\beqa
\label{survivaltransitionprobsdef}
P_{ee} & = & \frac{1}{2}\Tr(1 + \vec\sigma\cdot\vec e_3)\rho(t)) =
\frac{1}{2}(1 + \vec e_3\cdot\vec\rho)\,,\nonumber\\
P_{e\mu} & = & \frac{1}{2}\Tr(1 - \vec\sigma\cdot\vec e_3)\rho(t) =
\frac{1}{2}(1 - \vec e_3\cdot\vec\rho)\,,
\eeqa
can be computed using
\beq
\vec e_3\cdot\vec\rho = e^{-\gamma_\ell t} +
\left[e^{-\gamma_t t}\cos(ht) - e^{-\gamma_\ell t}\right]\sin^2 2\theta_m\,,
\eeq
which yields
\beq
\label{examplesolDterm}
\left.\begin{array}{c}P_{ee} \\ P_{e\mu}\end{array}\right\} =
\frac{1}{2} \pm \frac{1}{2} e^{-\gamma_\ell t} \pm
\frac{1}{2}\left[e^{-\gamma_t t}\cos(ht) -
  e^{-\gamma_\ell t}\right]\sin^2 2\theta_m\,.
\eeq
For $\gamma_{\ell,t} = 0$ they reduce to the standard
decoherence-free solutions
\beqa
\label{Probs}
P_{ee} & = & 1 - \sin^2 2\theta_m\sin^2(ht/2)\,,\nonumber\\
P_{e\mu} & = & \sin^2 2\theta_m\sin^2(ht/2)\,,
\eeqa
where $\sin^2 2\theta_m$ is given in \Eq{sin2thetam}.

We wish to make the following observation. The approximation we have made
by neglecting the mixing terms in \Eq{rhoperpexpansion}, amounts to take
\beq
\label{Dmodel}
\vec D = -\gamma_\ell\vec\rho_\ell - \gamma_t\vec\rho_t\,,
\eeq
in \Eq{eveqourcase}. This form of the equation has been used in previous works
that consider the decoherence effects, in which $\vec D$ is unknown and
treated at a phenomenological level\cite{footnote2}. In those
contexts the working assumption is that the decoherence terms are diagonal in
the basis of the effective mass eigenstates. In our notation this
translates to the statement that the $\vec D$ term
does not mix the $\vec\rho_\ell$ and $\vec\rho_t$ components of $\vec\rho$.
As we have seen, this is not strictly true for the $\vec D$ term
that we have calculated for the SM model. This is basically due to the
fact that the decoherence term that we have calculated is diagonal in
flavor space. Nevertheless, with the approximation we have
made above, we are able to make a correspondence with those
phenomenological treatments, with the bonus that we can give
a definite value for the $\gamma_{\ell,t}$ coefficients that
appear in \Eq{Dmodel} and parametrize the decoherece effects
as the example in \Eq{examplesolDterm} shows.
\section{Evaluation of integrals in various limiting cases}
\label{sec:evaluationintegrals}

For illustrative purposes and a guide to applications to realistic and/or
potentially important situations, here we evaluate explicitly the integrals
involved for some specific simple cases of the background conditions.

We assume that $f_{x} \ll 1$ so that we can set
$(1 -f_{x}(E_{p^\prime})) \rightarrow 1$. Then
\beqa
\label{I12example}
I^{(x)}_1 & = & \frac{2}{\omega_\kappa}\left(\frac{1}{2\pi}\right)^5
\int\frac{d^3p}{2E_{p}} f_x(E_p) J^{(2)}_1(p,k)\,,\nonumber\\
I^{(x)}_2 & = & \frac{2}{\omega_\kappa}\left(\frac{1}{2\pi}\right)^5
\int\frac{d^3p}{2E_{p}} f_{x}(E_p) J^{(2)}_2(p,k)\,,
\eeqa
where
\beqa
\label{J12def}
J^{(n)}_1 & = & \int\frac{d^3p^\prime}{2E_{p^\prime}}
\frac{d^3\kappa^\prime}{2\omega_{\kappa^\prime}}
\delta^{(4)}(p + k - p^\prime - k^\prime)(p\cdot k^\prime)^n\,,\nonumber\\
J^{(n)}_2 & = & \int\frac{d^3p^\prime}{2E_{p^\prime}}
\frac{d^3\kappa^\prime}{2\omega_{\kappa^\prime}}
\delta^{(4)}(p + k - p^\prime - k^\prime)(p\cdot k)^n\,.
\eeqa
For $I^{(x)}_0$ we use the following identity
%
%
which follows from momentum conservation,
\beq
(k - k^\prime + p)^2 = p^{\prime\,2} \Rightarrow
(k - k^\prime)^2 + 2p\cdot(k - k^\prime)  = 0\Rightarrow
k\cdot k^\prime = p\cdot(k - k^\prime)\,.
\eeq
Thus,
\beq
\label{I0example}
I^{(x)}_0 = \frac{2}{\omega_\kappa}\left(\frac{1}{2\pi}\right)^5
\int\frac{d^3p}{2E_{p}} f_x(E_p) (J^{(1)}_2(p,k) - J^{(1)}_1(p,k))\,,
\eeq
The integrals $J^{(2)}_{1,2}$ were denoted by $J_{1,2}$
in \xRef{nsnuphidecoherence}, and were evaluated there.
Imitating the procedure followed there, they can be
evaluated for any $n$, and in particular for $n = 1$.
The details are given in \Appendix{sec:integralsJ}.
Here we quote the results for particular cases
that can serve as a guide and benchmark when considering more general
situations. We consider separately the ultrarelativistic or a non-relativistic
fermion background, and specific limits of the thermal distributions.

\subsection{Ultrarelativistic background}

Specifically we assume that
\beq
\alpha_f, T,\omega_\kappa \gg m_f\,.
\eeq
In this case, as shown in \Appendix{sec:integralsJ}
\beqa
\label{J12nurlimit}
J^{(n)}_1 & = & \frac{\pi}{2(n + 1)}\omega^n_\kappa p^n(1 - \cos\theta_p)^n\,,
\nonumber\\
J^{(n)}_2 & = & (n + 1)J^{(n)}_1\,,
\eeqa
where $\theta_p$ is the angle between $\vec p$ and $\vec\kappa$,
and we have set $p = |\vec p|$. In particular,
\beqa
J^{(2)}_2 & = & 3 J^{(2)}_1\,,\nonumber\\
J^{(1)}_2 - J^{(1)}_1 & = & J^{(1)}_1\,.
\eeqa
Substituting these in \Eqs{I12example}{I0example}, and remembering
that $\omega_\kappa = \kappa$, then we obtain for this case,
\beqa
I^{(x)}_1 & = & \frac{\kappa}{36\pi^3}\int^\infty_0 dp p^3 f_x(p)\,,\nonumber\\
I^{(x)}_2 & = & 3I^{(x)}_1\nonumber\\
I^{(x)}_0 & = & \frac{1}{32\pi^3}\int^\infty_0 dp p^2 f_{x}(p)\,.
\eeqa
To carry out the integrals for $I^{(x)}_{0,1,2}$ we consider
separately the completely degenerate or the classical fermion
distribution.

\subsubsection{Completely degenerate background}

For a completely degenerate $x$ background ($x = f$ or $\bar f$)
putting $f_x = \theta(p_{Fx} - p)$, where $p_{Fx}$ is the Fermi momentum,
\beqa
I^{(x)}_1 & = & \frac{\kappa}{36\pi^3}\frac{p^4_{Fx}}{4}\,,\nonumber\\
I^{(x)}_2 & = & 3I^{(x)}_1\,,\nonumber\\
I^{(x)}_0 & = & \frac{1}{32\pi^3}\frac{p^3_{Fx}}{3}\,.
\eeqa
The Fermi momentum is given in terms of the number density $f_x$
of the background fermions by $p_{Fx} = (3\pi^2 n_x)^{\frac{4}{3}}$.

\subsubsection{Classical background}

Putting $f_x = e^{-\beta p}$, where $\beta$ is the
inverse temperature ($T$), gives
\beqa
\label{integralserclassical}
I^{(x)}_1 & = & \frac{\kappa T^4}{6\pi^3}\,,\nonumber\\
I^{(x)}_2 & = & 3I^{(x)}_1\,,\nonumber\\
I^{(x)}_0 & = & \frac{T^3}{16\pi^3}\,.
\eeqa

\subsection{Nonrelativistic background}

Here we assume that
\beq
m_f \gg T\,.
\eeq
We consider two situations separately, depending on whether
$\omega_\kappa \gg m_f$ or $\omega_\kappa \ll m_f$.

\subsubsection{$\omega_\kappa \gg m_f$}

In this case we obtain
\beqa
J^{(n)}_1 & = & \frac{\pi}{2(n + 1)}(m_f\omega_\kappa)^n\,,\nonumber\\
J^{(n)}_2 & = & (n + 1) J^{(n)}_1\,.
\eeqa
Then from \Eqs{I12example}{I0example},
\beqa
\label{integralsnrhigh}
I^{(x)}_1 & = & \frac{\kappa m_f n_x}{48\pi}\,,\nonumber\\
I^{(x)}_2 & = & 3I^{(x)}_1\,,\nonumber\\
I^{(x)}_0 & = & \frac{n_x}{32\pi}\,,
\eeqa
with
\beq
n_x = 2\int\frac{d^3p}{(2\pi)^3} f_x(E_p)\,.
\eeq

\subsubsection{$\omega_\kappa \ll m_f$}

In this case we obtain,
\beqa
J^{(2)}_2 = J^{(2)}_1 & = & \pi m_f\omega^3_\kappa\,,\nonumber\\
J^{(1)}_2 - J^{(1)}_1 & = & \frac{\pi\omega^3_\kappa}{m_f}\,,
\eeqa
and then from \Eqs{I12example}{I0example},
\beqa
\label{integralsnrlow}
I^{(x)}_1 & = & I^{(x)}_2 = \frac{\kappa^2 n_x}{8\pi}\,,\nonumber\\
I^{(x)}_0 & = & \frac{3}{4\pi}\frac{\kappa^3 n_x}{m^3_f}\,.
\eeqa

\section{Examples and Discussion}
\label{sec:discussion}

Here we use the results of the previous section to evaluate
$\gamma^{(W)}_e$, which drives the decoherence term
in \Eq{eveqLW}, in various environments of potential interest.
One important result is that the
formulas we have derived predict a well-defined and calculable
energy dependence of the decoherence terms
once the conditions of the environment are specified. This result
is in itself important in the context of recent studies that
have explored the possible energy dependence of the decoherence
terms, but from a phenomenological point of
view (e.g., Refs.\cite{lisi,farzan,Coloma:2018idr,Gomes:2020muc}).
Below we give the formulas for $\gamma^{(W)}_e$,
which enters in the evolution equation as indicated in \Eq{eveqLW},
but as already mentioned in \Section{subsec:decoherence},
for completeness we give the formulas for $\gamma^{(Z)}_f$ as well.

\subsection{Matter background}

As our first example we consider a normal matter background,
that is a medium consisting of non-relativistic electrons
and nucleons $N = n,p$) with no antiparticles. We consider three situations
separately, according to whether the neutrino energy is larger
or smaller than $m_e$ and the nucleon mass $m_N$.

\subsubsection{$\kappa > m_e,m_N$}
\label{subsubsec:gammacase1}

In this case we use \Eq{integralsnrhigh} for the electron and nucleon
backgrounds. Then from \Eqs{gamma}{gammabar}
\beqa
\label{gammacase1}
\gamma^{(W)}_e & = & \frac{\kappa}{16\pi}m_e n_e\,,\nonumber\\
\gamma^{(Z)}_f & = & \frac{\kappa}{16\pi}m_f n_f\left[(a_f - b_f)^2 +
  \frac{1}{3}(a_f + b_f)^2 - \frac{m_f}{2\kappa}(a^2_f - b^2_f)\right]\,,
\nonumber\\
\bar\gamma^{(W)}_e & = & \frac{\kappa}{48\pi}m_e n_e\,,\nonumber\\
\bar\gamma^{(Z)}_f & = & \frac{\kappa}{16\pi}m_f n_f
\left[\frac{1}{3}(a_f - b_f)^2 + (a_f + b_f)^2 -
\frac{m_f}{2\kappa}(a^2_f - b^2_f)\right]\,.
\eeqa
In some circumstances, it is possible that further approximations
are appropriate. For example, in the very high energy neutrino limit,
$\kappa \gg m_N,m_e$, then the last term in $\gamma^{(Z)}_f$ can
be neglected.

However, the distinguishing feature of this case is that
the $\gamma^{(W,Z)}_f$ factors, and whence
all the decoherence terms,
scale linearly with the neutrino energy ($\sim \kappa$).
%
%

\subsubsection{$m_N,m_e > \kappa$}
\label{subsubsec:gammacase2}

In this case we use \Eq{integralsnrlow} for the electron and nucleon
backgrounds. Then from \Eqs{gamma}{gammabar},
\beqa
\label{gammacase2}
\gamma^{(Z)}_f = \bar\gamma^{(Z)}_f & = & \frac{\kappa^2 n_f}{4\pi}
\left[a^2_f + b^2_f - \frac{3\kappa}{m_f}(a^2_f - b^2_f)\right]\,,\nonumber\\
\gamma^{(W)}_e = \bar\gamma^{(W)}_e & = & \frac{\kappa^2 n_e}{4\pi}\,.
\eeqa
In this case, in contrast to the previous one, the decoherence terms scale
as $\kappa^2$, and they are the same for neutrinos and antineutrinos.
%
%

\subsubsection{$m_N > \kappa > m_e$}
\label{subsubsec:gammacase3}

In this case we must use the formulas given in \Eq{gammacase1}
for the contribution due to the electron background, and \Eq{gammacase2}
for the nucleon contribution. That is (using $N$ to denote a nucleon
$n$ or $p$),
\beqa
\label{gammacase3}
\gamma^{(W)}_e & = & \frac{\kappa}{16\pi}m_e n_e\,,\nonumber\\
\gamma^{(Z)}_e & = & \frac{\kappa}{16\pi}m_e n_e\left[(a_e - b_e)^2 +
  \frac{1}{3}(a_e + b_e)^2 - \frac{m_e}{2\kappa}(a^2_e - b^2_e)\right]\,,
\nonumber\\
\gamma^{(Z)}_N & = & \frac{\kappa^2 n_N}{4\pi}
\left[a^2_N + b^2_N - \frac{3\kappa}{m_N}(a^2_N - b^2_N)\right]\,.
\eeqa
and
\beqa
\bar\gamma^{(W)}_e & = & \frac{\kappa}{48\pi}m_e n_e\,,\nonumber\\
\bar\gamma^{(Z)}_e & = & \frac{\kappa}{16\pi}m_e n_e
\left[\frac{1}{3}(a_e - b_e)^2 + (a_e + b_e)^2 -
\frac{m_e}{2\kappa}(a^2_e - b^2_e)\right]\,,\nonumber\\
\bar\gamma^{(Z)}_N & = & \gamma^{(Z)}_N\,.
\eeqa
Consequently, the $\kappa$ dependence can be more complicated
than both of the cases above, involving
a combination of terms that scale like $\kappa$ and
terms that scale like $\kappa^2$.
  
\subsection{Relativistic electron-positron background}

For illustrative and reference purposes we now consider
a classical background of electrons and positrons in the extremely
relativistic limit. Using \Eq{integralserclassical},
\beqa
\gamma^{(Z)}_e = \bar\gamma^{(Z)}_e & = &
\left[(a^2_e + b^2_e)\frac{4\kappa T^4}{3\pi^3} -
(a^2_e - b^2_e)\frac{m^2_e T^3}{8\pi^3}\right]\,,\nonumber\\
\gamma^{(W)}_e = \bar\gamma^{(W)}_e & = & \frac{2\kappa T^4}{3\pi^3}\,.
\eeqa
%

\subsection{Discussion}

By combining the formulas given above we can consider
other cases, for example, a background consisting of relativistic
electrons and positrons, superimposed on non-relativistic nuclear
matter. In general case, the dependence on $\kappa$ and/or $T$ is
not a single power law, as the examples above illustrate. Such
dependences are different depending  on the composition and conditions of
the background, and therefore in practical applications
it is necessary to specify the conditions of the background
medium in the context being considered.

For guidance let us consider two specific cases, which are
representative of the conditions that are relevant for long
baseline experiments. We consider the neutrino energy in two
different ranges and the neutrino oscillation parameters in the range
corresponding to atmospheric $\mu-\tau$ neutrino oscillations.
For this estimate, we take the same number density for electrons, protons
and neutrons and normalize it to $n_e = 10^{24}n_0$ cm$^{-3}$.

\begin{description}

\item[(i)] $m_N > \kappa > m_e$. This case is considered in
  \Section{subsubsec:gammacase3}. Here, we take neutrino energy
  $\kappa = 100\kappa_0$ MeV, which gives
  \beq
  \gamma^{(W)}_e = 7.8\times 10^{-24}\kappa_0 n_0\,\mbox{GeV}^5\,,
  \eeq
  In this case, the matter mixing angle is such that
  $\sin\,2\theta_m \simeq 1$ which gives
  \beq
    \label{gammaelltcasei}
  \gamma_t \simeq \frac{\gamma_\ell}{2}
  \simeq 2.1\times 10^{-33}\kappa_0 n_{0}\,\mbox{GeV}\,.
  \eeq
  In the case of active neutrinos with standard interactions
  only the $\gamma^{(W)}_e$ contributes to the decoherence terms.
  Since in non-standard cases $\gamma^{(Z)}_{N}$ can also contribute,
  for completeness we also quote the corresponding estimates,
  \beq
  \gamma^{(Z)}_{f} \simeq\left\{ 
  \begin{array}{ll}
    2.8\times 10^{-24}\kappa_0 n_0\,\mbox{GeV}^5 & \quad (f = e)\\
    2.4\times 10^{-21}\, \kappa^2_0 n_{0}\,\mbox{GeV}^5, & \quad (f = p)\\
    4.0\times 10^{-21}\,\kappa^2_0 n_{0}\,\mbox{GeV}^5 & \quad (f = n)
  \end{array}\right.
  \eeq
  In obtaining these values we have neglected the
  terms proportional to $m_e/\kappa$ and $\kappa/m_N$ in the
  formulas for $\gamma^{(Z)}_{f}$.
    
\item[(ii)] $\kappa > m_e,m_N$. This is the case considered in
  \Section{subsubsec:gammacase1}. For this case, we take the
  neutrino energy $\kappa = 100\kappa_0\,\mbox{GeV}$, which gives
  \beq 
  \gamma^{(W)}_e = 7.8\times 10^{-21}\kappa_0n_{0}\,\mbox{GeV}^5\,.
  \eeq
  The matter mixing angle for this case is such that $\sin^2 2\theta_m \simeq
  1.5\times 10^{-3}$, and therefore in this case
  $\gamma_l \ll \gamma_t$. Specifically,
  \beqa
  \label{gammaelltcaseii}
  \gamma_l &\simeq & 6.6\times 10^{-33}\, \kappa_0 n_{0}\,\mbox{GeV},\nonumber\\
  \gamma_t &\simeq & 4.3\times 10^{-30}\, \kappa_0 n_{0}\,\mbox{GeV}\,.
  \eeqa
  As we see, as the neutrino energy increases from 100 MeV to 100 GeV,
  in which case $\cos^2 2\theta_m$ goes from zero to unity,
  $\gamma_l $ goes from $\gamma_l \simeq 2\gamma_t$
  to $\gamma_l \ll \gamma_t$. Thus, for higher neutrino energies
  the main contribution to the decoherence terms comes 
  mainly from $\gamma_t$.

  As in the previous case, we quote the corresponding values for the 
  $ \gamma^{(Z)}_{f}$ terms,
  \beq
  \gamma^{(Z)}_{f} \simeq \left\{ 
  \begin{array}{l l}
    2.8\times 10^{-21}\kappa_0 n_{0}\,\mbox{GeV}^5, & \quad (f = e)\\
    8.0\times 10^{-18}\kappa_0 n_{0}\,\mbox{GeV}^5, & \quad (f = p)\\
    1.8\times 10^{-17}\kappa_0 n_{0}\,\mbox{GeV}^5, & \quad (f = n)
  \end{array} \right.
  \eeq
  where we have neglected the terms proportional to $m_f/\kappa$.

\end{description}

For reference we note that previous studies that have studied the effects
of the decoherence terms in the context of long baseline neutrino
oscillation from a purely phenomenological point of view
constrain the decoherence parameters corresponding to $\gamma_{\ell,t}$
to be less than $\sim 10^{-23}$ GeV, $10^{-24}$ GeV, depending
on the channel(see, e.g., \xRef{oliveira:hepph160308065}).
Comparing these values with our estimates in
\Eqs{gammaelltcasei}{gammaelltcaseii}
it seems that the SM decoherence terms are no consequence for the
long basseline experiments.

Being able to determine the value of these terms, as a result of a consistent
calculation, is useful because they serve as benchmark values against
which to compare contributions to decoherence from
other sources, for example from non-standard neutrino interactions,
and to assess the significance of deviations from
standard expectations with the decoherence terms not included.
In addition the possible applications of these results in other physical
contexts and environments should be kept in mind.
For example, for a background that is particle-antiparticle
symmetric, the leading contribution to the neutrino
effective potential $V_e$ is proportional to $m^{-4}_W$, and in such
environment the decoherent and coherent terms can be comparable.
\section{Conclusions and outlook}
\label{sec:conclusions}

In this work we have considered the effects of the
non-forward neutrino scattering processes on the propagation
of neutrinos in a matter (electron and nucleon) background.
Specifically, we calculated the contribution to the imaginary part of the
neutrino thermal self-energy arising from the non-forward neutrino
scattering processes in such backgrounds.
Since in this case the initial neutrino state is
depleted but does not actually disappear, we have argued that
such processes should be associated with decoherence effects.
More precisely, the non-forward scattering processes
produce a stochastic contribution to the evolution of the system that
cannot be described in terms of the coherent evolution of the state vector.
Following this view, we have given a precise
prescription to determine the jump operators, as used in the context
of the master or Lindblad equation, in terms of the results of
the calculation of the non-forward neutrino scattering
contribution to the imaginary part of the neutrino self-energy.
The main result is a well-defined formula for the jump operators,
expressed in terms of integrals over the background
matter fermion distribution functions and standard model couplings of the
neutrino with the electron and nucleons. For illustrative purposes
and guide to estimating the decoherence terms in situations 
of practical interest we gave explicit formulas for the decoherence terms
for different background conditions,
and pointed out some of the salient features in particular the
neutrino energy dependence. Our results indicate that
the effects of the decoherence terms are not appreciable in the context
of long baseline experiments. In any case, our results serve
as reference values to assess the significance of deviations from
standard expectations with the decoherence terms not included.
Their possible implications in other physical contexts should also
be kept in mind,  such as in particle-antiparticle
symmetric backgrounds in which case the decoherent and coherent terms
can be comparable.

The work of S. S. is partially supported by DGAPA-UNAM
(Mexico) Project No. IN103019.

\appendix
\section{Derivation of \Eq{SigmaW2}}
\label{sec:identityBCD}

We first prove the Fierz-like identity
\beq
\label{identityBCD}
\left(\Tr\gamma_\mu L B\gamma_\nu L C\right)\gamma^\mu L D\gamma^\nu L =
\left(\Tr\gamma_\mu L D\gamma_\nu L C\right)\gamma^\mu L B\gamma^\nu L\,,
\eeq
which is valid for any $4 \times 4$ matrices $B, C, D$. The proof is based
on another Fierz-like identity
\beq
\label{identityA}
\left(\Tr\gamma_\alpha L A\right)\gamma^\alpha L =
-\gamma_\alpha L A\gamma^\alpha L\,,
\eeq
which is valid for any $4\times 4$ matrix $A$. We write the
term in left-hand side of \Eq{identityBCD} in the form
\beq
\left(\Tr\gamma_\mu L B\gamma_\nu L C\right)\gamma^\mu L D\gamma^\nu L =
\left(\Tr\gamma_\mu L A\right)\gamma^\mu L D\gamma^\nu L\,,
\eeq
with
\beq
A = B\gamma_\nu L C\,.
\eeq
Applying \Eq{identityA}, we then get
\beq
\label{identityBCDpart1}
\left(\Tr\gamma_\mu L B\gamma_\nu L C\right)\gamma^\mu L D\gamma^\nu L =
-\gamma_\mu L B\gamma_\nu L C \gamma^\mu L D\gamma^\nu L\,.
\eeq
Now the term on the right hand side of this relation can be written
in the form
\beq
\gamma_\mu L B\gamma_\nu L C \gamma^\mu L D\gamma^\nu L =
\gamma_\mu L B \gamma_\nu L A^\prime\gamma^\nu L\,,
\eeq
with
\beq
A^\prime = C \gamma^\mu L D\,,
\eeq
and applying \Eq{identityA} again then yields
\beq
\label{identityBCDpart2}
\gamma_\mu L B\gamma_\nu L C \gamma^\mu L D\gamma^\nu L =
-\gamma_\mu L B \gamma_\nu L\left(\Tr\gamma^\nu L C\gamma^\mu LD\right) =
-\left(\Tr\gamma_\mu LD\gamma_\nu L C\right)\gamma^\mu L B \gamma^\nu L\,.
\eeq
Combining \Eqs{identityBCDpart1}{identityBCDpart2} leads to \Eq{identityBCD}.
Equation (\ref{SigmaW2}) follows from \Eq{SigmaW1} by applying the identity
in \Eq{identityBCD} with the identification
\beqa
B & = & iS^{(\nu_{Le})}_{12}(k^\prime)\,,\nonumber\\
C & = & iS^{(e)}_{21}(p)\,,\nonumber\\
D & = & iS^{(e)}_{12}(p^\prime)\,.
\eeqa
\section{Calculation of integrals $J^{(n)}_{1,2}$ in \Eq{J12def}}
\label{sec:integralsJ}

Since $J^{(n)}_{1,2}$ are a scalar integrals, we choose to do the integration
in the frame in which $p^\mu = (m_f,\vec 0)$ (the \emph{lab frame}). We
label the quantities in that frame with an asterisk,
$k^\mu = (\omega^\ast_\kappa,\vec\kappa^\ast)$ and similarly for
$k^{\prime\mu}$, and therefore
\beqa
J^{(n)}_1 & = & \int\frac{d^3\kappa^{\ast\prime}}{2\omega^\ast_{\kappa^\prime}}
\delta[(p + k - k^\prime)^2 - m^2_f]
\theta(m_f + \omega^\ast_\kappa - \omega^\ast_{\kappa^\prime})
(m_f\omega^\ast_{\kappa^\prime})^n\nonumber\\
& = & \int\frac{d^3\kappa^{\ast\prime}}{2\omega^\ast_{\kappa^\prime}}
\delta[-2\omega^\ast_\kappa \omega^\ast_{\kappa^\prime}
  (1 - \cos\theta^\ast_{\kappa^\prime}) +
2m_f(\omega^\ast_\kappa - \omega^\ast_{\kappa^\prime})]
\theta(m_f + \omega^\ast_\kappa - \omega^\ast_{\kappa^\prime})
(m_f\omega^\ast_{\kappa^\prime})^n\,,
\eeqa
where $\theta^\ast_{\kappa^\prime}$ is the angle between $\vec\kappa^\ast$
and $\vec\kappa^{\ast\prime}$. Carrying out with the integration over
$\cos\theta^\ast_{\kappa^\prime}$ first, with the help of the $\delta$ function,
yields
\beq
\cos\theta^\ast_{\kappa^\prime} = 1 -
\frac{m_f}{\omega^\ast_{\kappa}\omega^\ast_{\kappa^\prime}}
\left(\omega^\ast_{\kappa} - \omega^\ast_{\kappa^\prime}\right)\,,
\eeq
and
\beqa
\label{J1resultA}
J^{(n)}_1 & = &\frac{\pi m^n_f}{2\omega^\ast_\kappa}
\int^{\omega^{\ast\prime}_{max}}_{\omega^{\ast\prime}_{min}}
d\omega^{\ast}_{\kappa^\prime}\,
\omega^{\ast\,n}_{\kappa^\prime}\nonumber\\
& = & \frac{\pi m^n_f}{2(n + 1)\omega^\ast_\kappa}
\left(\omega^{\ast\prime\,n + 1}_{max} -
\omega^{\ast\prime\,n + 1}_{min}\right)\,,
\eeqa
where the requirement that $-1 \le \cos\theta^\ast_{\kappa^\prime} \le 1$
implies
\beqa
\label{omegaprimelabminmax}
\omega^{\ast\prime}_{min} & = &
\frac{m_f\omega^\ast_\kappa}{m_f + 2\omega^\ast_\kappa}\,,\nonumber\\
\omega^{\ast\prime}_{max} & = & \omega^\ast_\kappa\,.
\eeqa
For $J^{(n)}_2$ we proceed similarly, with the replacement
$p\cdot k^\prime \rightarrow p\cdot k = m_f\omega^\ast_\kappa$
in the integrand, and thus,
\beqa
\label{J2resultA}
J^{(n)}_2 & = & \frac{\pi m^n_f\omega^{\ast\,n - 1}_\kappa}{2}
\left(\omega^{\ast\prime}_{max} - \omega^{\ast\prime}_{min}\right)\,.
\eeqa

In order to use \Eqs{J1resultA}{J2resultA} in \Eqs{I12example}{I0example},
we express $\omega^{\ast\prime}_{min}$ and $\omega^{\ast\prime}_{max}$ in terms
of $E_p$ and $|\vec p|$ by means of the relation
\beq
\label{omegalab}
\omega^\ast_\kappa = \frac{1}{m_f}p\cdot k = \frac{\omega_\kappa E_p}{m_f}
(1 - v_p\cos\theta_p)\,,
\eeq
with $v_p = |\vec p|/E_p$. This allows the angular
integration in \Eq{I12example}
to be carried out in straightforward fashion, leaving only the integration
over $E_p$, which depends on the distribution function, to be performed.
As usual we can consider special cases for illustrative purposes.

\subsection{Ultrarelativistic background}

Specifically we assume that
\beq
\alpha_f, T,\omega_\kappa \gg m_f\,.
\eeq
In this case,
\beq
\label{wprimeastminzero}
\omega^{\ast\prime}_{min} = 0\,,
\eeq
and therefore
\beq
J^{(n)}_1 = \frac{\pi}{2(n + 1)}(m_f\omega^\ast_\kappa)^n \rightarrow
\frac{\pi}{2(n + 1)}\omega^n_\kappa p^n(1 - \cos\theta_p)^n\,,
\eeq
where we have set $p = |\vec p|$. Similarly,
\beq
J^{(n)}_2 = \frac{\pi}{2}(m_f\omega^\ast_\kappa)^n = (n + 1)J^{(n)}_1\,.
\eeq
In particular,
\beq
J^{(2)}_2 = 3 J^{(2)}_1\,,
\eeq
and
\beq
J^{(1)}_2 - J^{(1)}_1 = J^{(1)}_1\,,
\eeq
Thus from \Eqs{I12example}{I0example},
\beqa
I^{(x)}_1 & = &
\frac{2}{\omega_\kappa}\left(\frac{1}{2\pi}\right)^5
\frac{\pi^2\omega^2_\kappa}{6}
\frac{8}{3}\int^\infty_0 dp p^3 f_f(p)\nonumber\\
& = & \frac{\kappa}{36\pi^3}\int^\infty_0 dp p^3 f_x(p)\,,\nonumber\\
I^{(x)}_2 & = & 3I^{(x)}_1\nonumber\\
& = & \frac{\kappa}{12\pi^3}\int^\infty_0 dp p^3 f_{x}(p)\,,\nonumber\\
I^{(x)}_0 & = & \frac{1}{32\pi^3}\int^\infty_0 dp p^2 f_{x}(p)\,,
\eeqa
remembering that $\omega_\kappa = \kappa$.

\subsubsection{Completely degenerate background}

For a completely degenerate $x$ background ($x = f$ or $\bar f$)
putting $f_x = \theta(p_{Fx} - p)$, where $p_{Fx}$ is the Fermi momentum,
\beqa
I^{(x)}_1 & = & \frac{\kappa}{36\pi^3}\frac{p^4_{Fx}}{4}\,,\nonumber\\
I^{(x)}_2 & = & 3I^{(x)}_1 =
\frac{\kappa}{12\pi^3}\frac{p^4_{Fx}}{4}\,,\nonumber\\
I^{(x)}_0 & = & \frac{1}{32\pi^3}\frac{p^3_{Fx}}{3}\,.
\eeqa
The Fermi momentum is given in terms of the number density $f_x$
of the background fermions by $p_{Fx} = (3\pi^2 n_x)^{\frac{4}{3}}$.

\subsubsection{Classical background}

Putting $f_x = e^{-\beta p}$, where $\beta$ is the
inverse temperature ($T$), gives
\beqa
I^{(x)}_1 & = & \frac{6\kappa}{36\pi^3 \beta^4} = \frac{\kappa T^4}{6\pi^3}\,,
\nonumber\\
I^{(x)}_2 & = & 3I^{(x)}_1 = \frac{\kappa T^4}{2\pi^3}\,,\nonumber\\
I^{(x)}_0 & = & \frac{T^3}{16\pi^3}\,.
\eeqa

\subsection{Nonrelativistic background}

Here we assume that
\beq
m_f \gg T\,.
\eeq
From \Eq{omegalab},
\beq
\label{omegalabnr}
\omega^\ast_\kappa = \omega_\kappa\,.
\eeq
We consider two situations separately, depending on whether
$\omega_\kappa \gg m_f$ or $\omega_\kappa \ll m_f$.

\subsubsection{$\omega_\kappa \gg m_f$}

In this case we have \Eq{wprimeastminzero} once again. Thus,
\beq
J^{(n)}_1 \rightarrow \frac{\pi}{2(n + 1)}(m_f\omega_\kappa)^n\,,
\eeq
and similarly we get
\beq
J^{(n)}_2 = (n + 1) J^{(n)}_1\,,
\eeq
as in the ultrarelativistic case. Then from \Eqs{I12example}{I0example},
\beqa
I^{(x)}_1 & = & \frac{\kappa m_f n_x}{48\pi}\,,\nonumber\\
I^{(x)}_2 & = & 3I^{(x)}_1 = \frac{\kappa m_f n_x}{16\pi}\,,\nonumber\\
I^{(x)}_0 & = & \frac{n_x}{32\pi}\,,
\eeqa
with
\beq
n_x = 2\int\frac{d^3p}{(2\pi)^3} f_x(E_p)\,.
\eeq

\subsubsection{$\omega \ll m_f$}

From \Eq{omegaprimelabminmax}, we write
\beq
\omega^{\ast\prime}_{min} = \omega^\ast_\kappa\left(\frac{1}{1 + x}\right)\,,
\eeq
where
\beq
x = \frac{2\omega^\ast_\kappa}{m_f}\,.
\eeq
Therefore, from \Eqs{J1resultA}{J2resultA},
\beqa
J^{(n)}_{1} & = & \frac{\pi}{2}(m_f\omega^\ast_\kappa)^n f^{(n)}_{1}\,,
\nonumber\\
J^{(n)}_{2} & = & \frac{\pi}{2}(m_f\omega^\ast_\kappa)^n f_{2}\,,
\eeqa
where
\beqa
f^{(n)}_1 & = & \frac{1}{n + 1}
\left[1 - \left(\frac{1}{1 + x}\right)^{n + 1}\right]\,,\nonumber\\
f_2 & = & \frac{x}{1 + x}\,.
\eeqa
Thus, remembering \Eq{omegalabnr}, we then have
\beqa
J^{(2)}_1 & = & \frac{\pi}{2}(m_f\omega_\kappa)^2 \frac{2\omega_\kappa}{m_f}\,,
\nonumber\\
J^{(2)}_2 & = & J^{(2)}_1\,,\nonumber\\
J^{(1)}_2 - J^{(1)}_1 & = & \frac{\pi}{4}m_f\omega_\kappa
\left(\frac{2\omega_\kappa}{m_f}\right)^2\,,
\eeqa
to the leading order in $x$. Then from \Eqs{I12example}{I0example},
\beqa
I^{(x)}_1 & = & I^{(x)}_2 = \frac{\kappa^2 n_x}{8\pi}\,,\nonumber\\
I^{(x)}_0 & = & \frac{3}{4\pi}\frac{\kappa^3 n_x}{m^3_f}\,.
\eeqa

\subsection*{Details of Eqs. (B.26-III) and (B.27-II)}

From the definition of the $f$'s, I find
\beqa
f_2 - f^{(1)}_1 & = & \frac{x}{1 + x} - \frac{1}{2}\left[1 -
  \left(\frac{1}{1 + x}\right)^2\right]\nonumber\\
& = & \frac{1}{2}\frac{x^2}{(1 + x)^2}\,,
\eeqa
exactly. To leading order, then
\beq
f_2 - f^{(1)}_1 \simeq \frac{1}{2}x^2\,,
\eeq
and
\beqa
J^{(1)}_2 - J^{(1)}_1 & = & \frac{\pi}{2}m_f \omega^\ast_\kappa
\left(\frac{1}{2}x^2\right)\nonumber\\
& \rightarrow & \frac{\pi}{4}m_f \omega_\kappa
\left(\frac{2\omega_\kappa}{m_f}\right)^2
\eeqa


\begin{thebibliography}{99}
%
\bibitem{ftft:reviews}
  See for example, N.~P.~Landsman and C.~G.~van Weert,
  \emph{Real and Imaginary Time Field Theory at Finite Temperature and Density},
  Phys.\ Rept.\  {\bf 145}, 141 (1987);
  %
  J.~I.~Kapusta,
 \emph{Finite Temperature Field Theory},
 (Cambridge University Press,Cambridge, 1989);
  %
  A.~K.~Das,
\emph{Finite Temperature Field Theory},
  (World Scientific Singapore, 1997);
  %
  M.~L.~Bellac,
  \emph{Thermal Field Theory}, (Cambridge University Press, Cambridge, 2011).

\bibitem{Nieves:2018vxl}
    For a recent application and references to previous works
    along these lines see, for example,
    J.\ F.\ Nieves and S.\ Sahu,
    \emph{Neutrino effective potential in a fermion and scalar background},
    Phys. Rev. D 98, 063003 (2018) [arXiv:1808.01629].

\bibitem{nsnuphidamp}
    J.\ F.\ Nieves and S.\ Sahu,
    \emph{Neutrino damping in a fermion and scalar background},
    Phys. Rev. D 99, 095013 (2019)
    [arXiv:1812.05672]

\bibitem{nsnuphidecoherence}
    J.\ F.\ Nieves and S.\ Sahu,
    \emph{Neutrino decoherence in a fermion and scalar background},
    Phys. Rev. D 100, 115049 (2019)
    [arXiv:1909.11271]

\bibitem{Daley:2014fha} 
  A.~J.~Daley,
\emph{Quantum trajectories and open many-body quantum systems},
  Adv.\ Phys.\  {\bf 63}, no. 2, 77 (2014)
  [arXiv:1405.6694].

\bibitem{Weinberg:2011jg} 
  S.~Weinberg,
 \emph{Collapse of the State Vector},
  Phys.\ Rev.\ A {\bf 85}, 062116 (2012)
  [arXiv:1109.6462].

\bibitem{pearle}
  P. Pearle,
  \emph{Simple derivation of the Lindblad equation},
  Eur. J. Phys. 33, 805 (2012),
  [arXiv:1204.2016].

\bibitem{Plenio:1997ep} 
  M.~B.~Plenio and P.~L.~Knight,
\emph{The Quantum jump approach to dissipative dynamics in quantum optics},
  Rev.\ Mod.\ Phys.\  {\bf 70}, 101 (1998)
  [arxiv:quant-ph/9702007].

\bibitem{Lieu:2019cev} 
  S.~Lieu,
  \emph{Non-Hermitian Majorana modes protect degenerate steady states},
  Phys.\ Rev.\ B {\bf 100}, 085110 (2019)
  [arXiv:1904.07481]

%
%
\bibitem{lisi}
    E. Lisi, A. Marrone, and D. Montanino,
    \emph{Probing Possible Decoherence Effects in Atmospheric
    Neutrino Oscillations}, PRL 85, 1166 (2000).

\bibitem{farzan}
    Yasaman Farzan, Thomas Schwetz, Alexei Yu Smirnov,
    \emph{Reconciling results of LSND, MiniBooNE and other
    experiments with soft decoherence},
    JHEP 0807, 067 (2008).
    [arxiv:0805.2098]

%
%
\bibitem{oliveira:hepph160308065}
  R.~L.~N.~Oliveira,
  \emph{Dissipative Effect in Long Baseline Neutrino Experiments},
  Eur. Phys. J. C 76, 417 (2016)
  [arxiv:1603.08065]

\bibitem{Oliveira:2014jsa} 
  M.~M.~Guzzo, P.~C.~de Holanda and R.~L.~N.~Oliveira,
 \emph{Quantum Dissipation in a Neutrino System Propagating
               in Vacuum and in Matter},
  Nucl.\ Phys.\ B {\bf 908}, 408 (2016)
  [arXiv:1408.0823].

\bibitem{Carpio:2017nui} 
  J.~A.~Carpio, E.~Massoni and A.~M.~Gago,
 \emph{Revisiting quantum decoherence for neutrino oscillations in matter
                  with constant density},
  Phys.\ Rev.\ D {\bf 97}, no. 11, 115017 (2018)
  [arXiv:1711.03680].

%
%
\bibitem{Coloma:2018idr}
  P.~Coloma, J.~Lopez-Pavon, I.~Martinez-Soler and H.~Nunokawa,
  \emph{Decoherence in Neutrino Propagation Through Matter,
                    and Bounds from IceCube/DeepCore},
  Eur.\ Phys.\ J.\ C {\bf 78}, no. 8, 614 (2018)
  [arXiv:1803.04438]

%
%
\bibitem{Gomes:2018inp}
  G.~Balieiro Gomes, D.~V.~Forero, M.~M.~Guzzo, P.~C.~De Holanda and
  R.~L.~N.~Oliveira,
  \emph{Quantum Decoherence Effects in Neutrino Oscillations at DUNE},
  Phys.\ Rev.\ D {\bf 100}, no. 5, 055023 (2019)
  [arXiv:1805.09818]

%
%
\bibitem{Coelho:2017byq}
  J.~A.~B.~Coelho and W.~A.~Mann,
   \emph{Decoherence, matter effect, and neutrino hierarchy signature in
   long baseline experiments},
  Phys.\ Rev.\ D {\bf 96}, no. 9, 093009 (2017).
  [arXiv:1708.05495].

\bibitem{Gomes:2020muc}
  A.~L.~G.~Gomes, R.~A.~Gomes and O.~L.~G.~Peres,
  \emph{Quantum decoherence and relaxation in neutrinos using
  long-baseline data},
  [arXiv:2001.09250]

%
%
\bibitem{Fogli:2007tx} 
  See, for example,
  G.~L.~Fogli, E.~Lisi, A.~Marrone, D.~Montanino and A.~Palazzo,
 \emph{Probing non-standard decoherence effects with
       solar and KamLAND neutrinos},
  Phys.\ Rev.\ D {\bf 76}, 033006 (2007)
  [arXiv:0704.2568],
  and references therein.

%
%
\bibitem{Capolupo:2018hrp}
  A.~Capolupo, S.~M.~Giampaolo and G.~Lambiase,
\emph{Decoherence in neutrino oscillations, neutrino nature and CPT violation},
  Phys.\ Lett.\ B {\bf 792}, 298 (2019)
  [arXiv:1807.07823].

\bibitem{Buoninfante:2020iyr}
  L.~Buoninfante, A.~Capolupo, S.~M.~Giampaolo and G.~Lambiase,
 \emph{Revealing neutrino nature and $CPT$ violation with decoherence effects},
  [arXiv:2001.07580]

\bibitem{footnote1}
   Strictly speaking this is correct in the massless neutrino limit,
   which in practice is a valid approximation in the limit that the neutrino
   mass can be neglected in the calculation of the relevant diagrams.

\bibitem{footnote2}
   $\gamma_{t,\ell}$ correspond to the the parameters named $\Gamma_{1,2}$ in
   \xRef{oliveira:hepph160308065}, for example.
\end{thebibliography}

\end{document}